\documentclass[%
reprint,
superscriptaddress,
amsmath,amssymb,
aps,
prb
floatfix,
showpacs,
]{revtex4-2}

\usepackage{graphicx}
\usepackage{dcolumn}
\usepackage{bm}
\usepackage{hyperref}
\usepackage{arydshln}
\usepackage{ulem}

\usepackage[english]{babel}
\usepackage[utf8]{inputenc}
\usepackage{amssymb,amsfonts,amsmath,mathtools,mathrsfs}
\usepackage{relsize}
\usepackage{mdframed}
\usepackage[margin=0.6in]{geometry}
\usepackage{enumitem}
\usepackage{graphicx}
\usepackage{float}
\usepackage{url,hyperref}
\usepackage[table,usenames,dvipsnames]{xcolor}
\usepackage{gensymb}
\usepackage{grffile}

\hypersetup{
	colorlinks=true,
	linkcolor=Blue,
	filecolor=magenta,      
	urlcolor=Blue,
	citecolor=Blue,
}
\usepackage[bottom]{footmisc}
\usepackage[capitalize]{cleveref}
\usepackage{booktabs}
\usepackage{textcomp}
\usepackage{braket} 

\setcitestyle{line}

\usepackage{amsmath,amssymb,wasysym}

\makeatletter
\def\@fnsymbol#1{\ensuremath{\ifcase#1\or *\or \dagger\or \ddagger\or
   \mathsection\or \mathparagraph\or \|\or **\or \dagger\dagger
   \or \ddagger\ddagger \else\@ctrerr\fi}}
    \makeatother

\begin{document}

\title{Disorder-driven competing magnetic interactions and glassy magnetic behavior in quaternary Heusler alloy FeRuMnGe}

\author{Manikantha Panda}
\address{Department of Physics, National Institute of Technology Andhra Pradesh, Tadepalligudem 534101, India}
\author{Prabuddha Kant Mishra}
\address{Department of Chemistry, Indian Institute of
Technology Delhi, New Delhi 110016, India}
\author{Sonali S. Pradhan}
\address{Department of Physics, Indian Institute of Technology Hyderabad, Kandi - 502285, Sangareddy, Telangana, India.}
\author{Aarti Gautam}
\address{Department of Chemistry, Indian Institute of
Technology Delhi, New Delhi 110016, India}
\author{Bhagyashree Pol}
\address{UGC-DAE Consortium for Scientific Research, Mumbai Centre, BARC Campus, Mumbai 400085, India}
\author{P D Babu}
\address{UGC-DAE Consortium for Scientific Research, Mumbai Centre, BARC Campus, Mumbai 400085, India}

\author{Ashok Kumar Ganguli}
\address{Department of Chemistry, Indian Institute of
Technology Delhi, New Delhi 110016, India}
\address{Department of Chemical Sciences, Indian Institute of Science Education and Research, Berhampur, Odisha-760003, India}
\author{V. Kanchana}
\email{kanchana@phy.iith.ac.in}
\address{Department of Physics, Indian Institute of Technology Hyderabad, Kandi - 502285, Sangareddy, Telangana, India.}

\author{Tapas Paramanik} \email[E-mail: ]{tapas.phys@nitandhra.ac.in}
\address{Department of Physics, National Institute of Technology Andhra Pradesh, Tadepalligudem 534101, India}

\begin{abstract}
In this combined experimental and theoretical study, we investigate the role of disorder in governing the magnetic ground state of the quaternary Heusler alloy FeRuMnGe. In the FeRuMnZ (Z = Ga, Si) series, chemical substitution modifies atomic ordering and electronic structure, resulting in distinct magnetic ground states. Motivated by this, we extend the series to FeRuMnGe. X-ray diffraction reveals B2-type antisite disorder, where Fe–Ru and Mn–Ge intermix. Theoretical calculations show that such disorder modifies exchange interactions, leading to competing ferromagnetic and antiferromagnetic couplings, and drives the system from half-metallic to metallic. Magnetic measurements reveal competing interactions, giving rise to a cluster-glass state coexisting with long-range magnetic order. The absence of a thermodynamic signature at $T_f$, together with ac susceptibility and relaxation measurements, supports the presence of short-range magnetic interactions and a glassy magnetic state. The compound exhibits an enhanced magnetic response below $\sim 161$ K and a maximum magnetization of $\sim 1.64~\mu_B$/f.u. at 5 T.
Overall, this work establishes a direct correlation between antisite disorder, competing exchange interactions, and glassy magnetism in quaternary Heusler alloys. Combined experimental results and theoretical calculations reveal that disorder drives the system from an AFM-dominated state in FeRuMnSi to an FM-dominated state in FeRuMnGe, providing deeper insight into the role of extent of disorder in governing the magnetic properties of QHAs.

\end{abstract}

\date{\today}


\maketitle

\section{Introduction}

In condensed matter physics, materials exhibiting an interplay between magnetic and transport properties, along with a tunable electronic structure, are highly desirable because the coupling among spin, charge, and electronic degrees of freedom gives rise to novel quantum phenomena and multifunctional properties.  
Heusler alloys provide an excellent platform for tailoring electronic and magnetic behavior through controlled variation of chemical composition and atomic arrangement \cite{graf2010heusler, Felser2011, Graf2016}. Because of their multiple crystallographic sublattices and adjustable site occupancy, quaternary Heusler alloys (QHAs) provide an edge to tune their physical properties \cite{Neibecker2017, Bainsla2016}. A variety of emergent phenomena, including half-metallicity \cite{CoRuFeSiAHE, kundu2017new}, spin-gapless semiconducting behavior \cite{CoFeMnSi_SGS}, unusual transport properties \cite{PANDA2026187369}, high spin polarization \cite{LK2014}, etc., are made possible by this structural flexibility in QHAs, which allows precise control over band structure and magnetic interactions. Because of these attributes, Heusler systems have attracted significant interest amongst the condensed matter physics for both fundamental research and potential applications in energy-related applications and magnetic sensing \cite{Graf2016, hirohata2022heusler}.

\begin{figure*}
\begin{center}
    \includegraphics[width=1\linewidth]{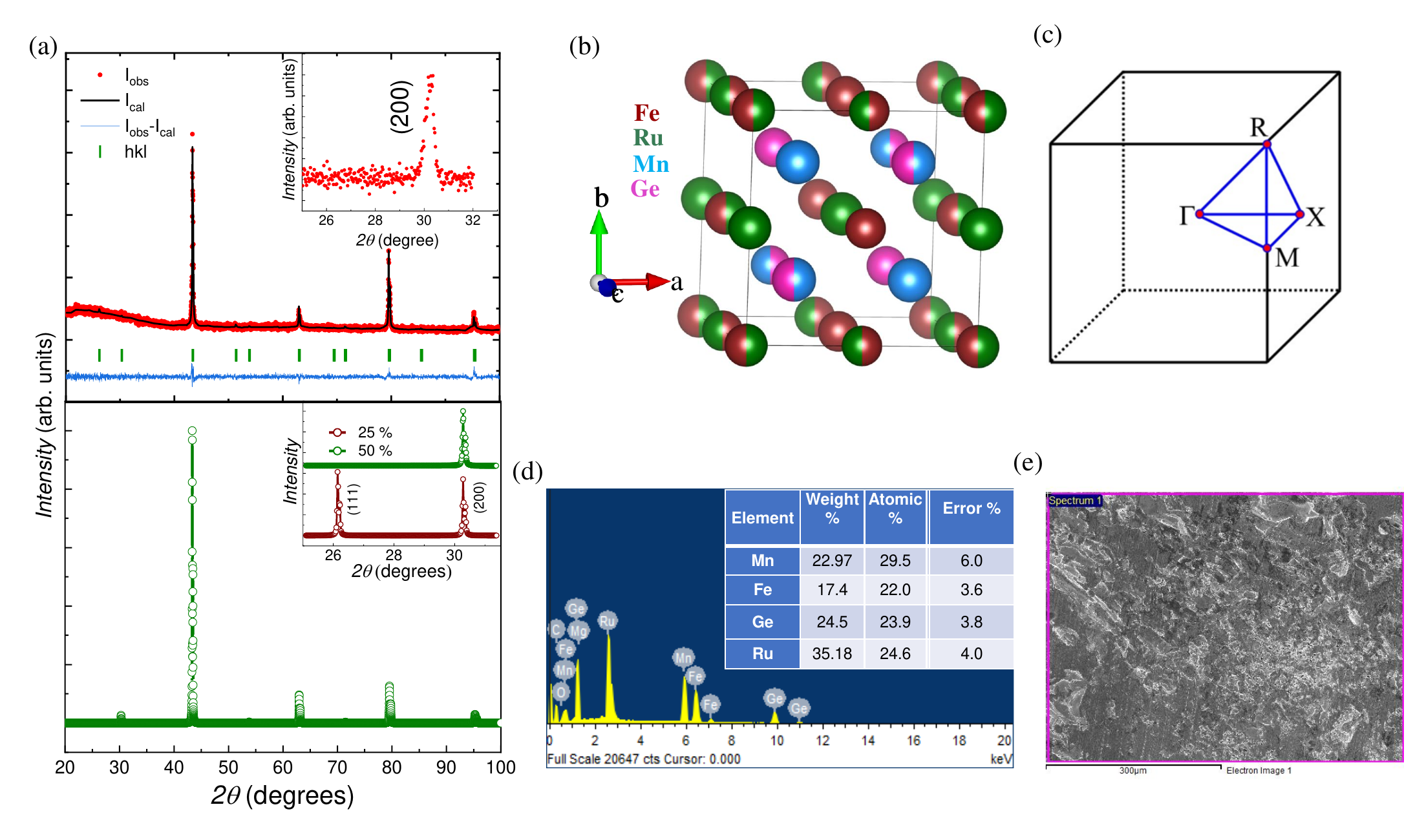}
    \caption{ (Color online) (a) Upper panel: Rietveld refinement of the room-temperature powder X-ray diffraction pattern of polycrystalline FeRuMnGe. The vertical tick marks indicate the allowed Bragg reflections, while the blue line represents the difference between the observed and fitted profiles. The inset shows the enlarged view of the (200) reflection peak. Lower panel: Simulated X-ray diffraction patterns of FeRuMnGe generated using VESTA software for 25\% and 50\% B2 disorder. The inset shows the magnified view of the (111) and (200) superlattice reflections for 25\% and 50\% B2 disorder. (b) Disordered (50\% B2) crystal structure of FeRuMnGe. (c) Corresponding Brillouin zone. (d) Energy-dispersive X-ray (EDX) analysis confirming the elemental composition of FeRuMnGe. (e) FE-SEM image of FeRuMnGe.}
    \label{fig1}
\end{center}
\end{figure*}


Despite their structural simplicity, Heusler alloys are highly susceptible to anti-site disorder due to the comparable electronegativity and atomic size of the constituent elements \cite{FeRuMnGa, graf2010heusler, Felser2011}. These systems are known to have a significant impact on their electronic structure due to anti-site disorder, which is exchange of atomic positions between various sublattices. From a materials design perspective, antisite disorder offers an effective means to tailor the electronic structure and magnetic interactions in Heusler alloys \cite{graf2010heusler}. This causes a redistribution of electronic states without significantly changing the metallic ground state by changing the density of states close to the Fermi level and the hybridization between constituent atoms \cite{ENAMULLAH20181055, VENKATESH2013417}. In addition, antisite disorder perturbs the regular arrangement of magnetic atoms, modifying exchange interactions in a way that suppresses long-range magnetic order and promotes competing interactions \cite{graf2010heusler,Sasioglu2005, GHOSH2025182719}.
 The coexistence of ferromagnetic and antiferromagnetic interactions leads to magnetic frustration, which can give rise to glassy states such as spin glass, cluster glass, or reentrant glass behavior in the material. These effects are typically manifested through slow relaxation dynamics and non-equilibrium magnetic behavior in QHAs \cite{FeRuMnGa,PANDA2026187369, NiRuMnSn}.
In metallic systems, competitive exchange interactions resulting from indirect exchange between localized moments control the magnetic ground state. In this regard, the Ruderman–Kittel–Kasuya–Yosida (RKKY) interaction provides a natural framework, as its oscillatory nature allows both ferromagnetic and antiferromagnetic interactions to coexist \cite{Ruderman1954}.  

From a chemical perspective, substitution of the main block element can significantly influence the degree of atomic disorder in Heusler alloys \cite{ZHANG201386, Varaprasad_2010}. In this context, the FeRuMnZ (Z = Ga, Si) series is known to exhibit a wide range of magnetic properties \cite{PANDA2026187369,FeRuMnGa}, where reduced magnetic moments and contrasting magnetic ground states highlight the sensitivity of magnetic behavior to subtle changes in electronic structure and chemical composition. Such sensitivity is often associated with competing magnetic interactions arising from disorder. Here, we investigate the FeRuMnGe compound, which exhibits a notable enhancement in the magnetic response compared to FeRuMnZ (Z = Ga, Si). In particular, the system exhibits cluster-glass behavior, corroborated by dc magnetization, ac susceptibility, relaxation, and heat capacity measurements. 
While glassy magnetic behavior is commonly associated with competing interactions, our approach provides a direct fundamental understanding of its origin by demonstrating that antisite disorder modifies the exchange interactions, giving rise to such competing magnetic behavior in this system. Although structural disorder has been identified in the closely related FeRuMnGa compound \cite{FeRuMnGa}, the present work establishes its decisive role in driving these interactions, thereby providing new insight into the emergence of complex magnetic behavior.

\begin{figure*}
    \centering
    \includegraphics[width=0.84\linewidth]{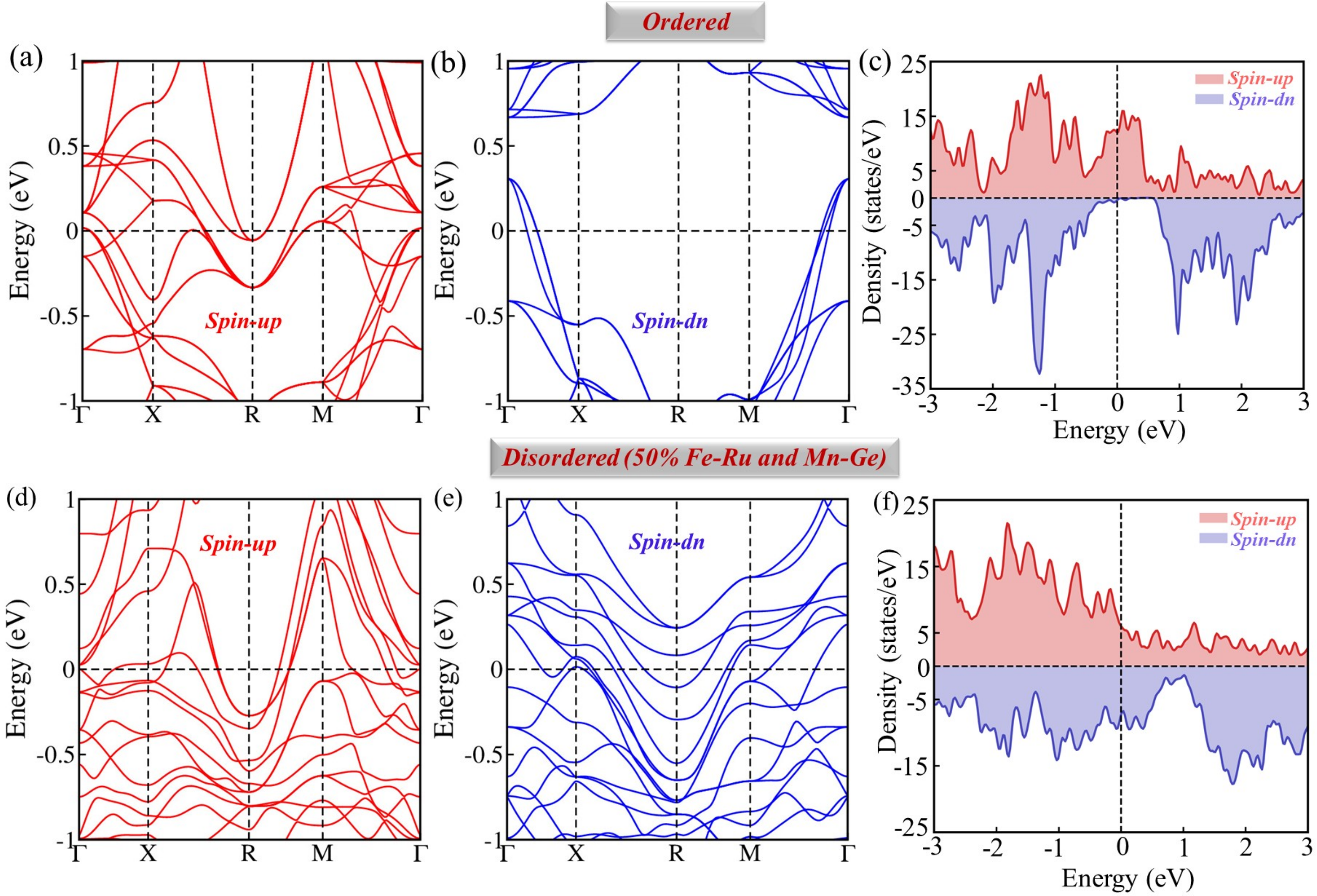}
    \caption{(a–c) Electronic band structure and density of states (DOS) of the ordered phase; (d–f) corresponding results for the Fe–Ru and Mn–Ge antisite-disordered phase. The spin-up and spin-down channels are represented by red and blue colors, respectively.}
    \label{ba}
\end{figure*}
\section{Experimental and Computational Details}

Polycrystalline samples of FeRuMnGe were prepared with the standard arc-melting method within a high-purity argon atmosphere. Stoichiometric quantities of high-purity elemental elements ($>$99.9\%, Alfa Aesar, Thermo Scientific) in the form of small pieces were used as initial materials. To compensate for the potential loss of Mn, attributed to its volatile characteristics during melting, an extra 5\% Mn was included. The resultant ingot underwent remelting five times, with inversion following each melting process, to guarantee compositional uniformity. Thereafter, the as-cast ingot was encapsulated in a quartz tube under vacuum and annealed at 1073 K for three and five days to enhance atomic ordering, followed by rapid quenching in ice-cold water. The resultant final sample demonstrated minimal weight reduction (less than 0.5\%).
Powder X-ray diffraction (PXRD) analyses were conducted utilizing a Malvern Panalytical diffractometer employing Cu-K$\alpha$ radiation ($\lambda$ = 1.5406 \AA). Rietveld refinement of the room-temperature PXRD data, executed with the FULLPROF suite \cite{FULLPROF}, validated the existence of a single-phase sample and facilitated the elucidation of its crystal structure.
Magnetization assessments were conducted utilizing a vibrating sample magnetometer (VSM) at magnetic fields reaching H = $\pm$ 5 T across a temperature spectrum of 2 - 300 K. The ac-susceptibility assessments were conducted with a magnetic property measurement system (PPMS, Quantum Design) over several excitation frequencies within the temperature spectrum of 50 - 85 K.

Structural optimization was performed utilizing the Vienna \textit{ab initio} simulation package (VASP) \cite{kresse1993ab, perdew1996generalized} under the auspices of density functional theory (DFT). The exchange-correlation effects were addressed utilizing the Perdew--Burke--Ernzerhof (PBE) functional \cite{perdew1996generalized} in accordance with the generalized gradient approximation (GGA). A plane-wave cutoff energy of 600 eV was utilized for all computations, and the total energy convergence threshold was established at \(10^{-8}\) eV. A $16\times16\times16$ k-point grid was utilized to sample the irreducible Brillouin zone (BZ) via the Monkhorst--Pack method \cite{monkhorst1976special}. Magnetic exchange interactions were calculated via a Wannier-based tight-binding method executed in the TB2J software package \cite{He2021}. In the framework of Green's function, the exchange parameters, especially those between Fe and Mn atoms, were derived by considering infinitesimal spin rotations as perturbative influences. The resultant couplings were employed to examine the spin dynamics.


\section{Results and Discussion}

\subsection{Structural and compositional analysis}

\hyperref[fig1]{Fig.~1(a)} shows the room-temperature PXRD pattern of the polycrystalline FeRuMnGe sample along with the Rietveld refinement profile, Bragg peak positions, and the corresponding difference curve. The excellent agreement between the observed and calculated patterns confirms the formation of a single-phase compound. The refinement reveals that FeRuMnGe crystallizes in the LiMgPdSn-type structure with space group $F\overline{4}3m$ (No.~216) and lattice parameter $\sim 5.898 (5)$~\r{A}. The atomic arrangement corresponds to Fe occupying the $4a$ $(0,0,0)$ site, Ru $4b$ $(\tfrac{1}{2},\tfrac{1}{2},\tfrac{1}{2})$ site, Mn $4c$ $(\tfrac{1}{4},\tfrac{1}{4},\tfrac{1}{4})$ site, and Ge $4d$ $(\tfrac{3}{4},\tfrac{3}{4},\tfrac{3}{4})$ site. This atomic configuration is consistent with that reported for FeRuMnSi \cite{PANDA2026187369} and is further supported by our theoretical calculations (as discussed in the later section).

The degree of atomic ordering in QHAs is commonly inferred from the superlattice reflections (111) and (200) in XRD data. For the present atomic configuration, the structure factor can be written as
\begin{align}
F_{hkl} &= F_{fcc}\Big[ f_{\text{Fe}} 
+ f_{\text{Ru}} e^{\pi i(h+k+l)} \notag \\
&\quad + f_{\text{Mn}} e^{\tfrac{\pi i}{2}(h+k+l)} 
+ f_{\text{Ge}} e^{-\tfrac{\pi i}{2}(h+k+l)} \Big],
\end{align}

where $F_{fcc}=4$ for unmixed $(h,k,l)$ reflections and zero otherwise, and $f_{\text{Fe}}$, $f_{\text{Ru}}$, $f_{\text{Mn}}$, and $f_{\text{Ge}}$ are the atomic scattering factors of the respective elements.

Accordingly, the superlattice reflections are given by

\begin{equation}
F_{111} = 4\left[(f_{\text{Fe}} - f_{\text{Ru}}) + i(f_{\text{Ge}} - f_{\text{Mn}})\right],
\end{equation}

\begin{equation}
F_{200} = 4\left[(f_{\text{Fe}} + f_{\text{Ru}}) - (f_{\text{Mn}} + f_{\text{Ge}})\right].
\end{equation}

These expressions indicate that mixing between Fe and Ru sites or between Mn and Ge sites can suppress the corresponding superlattice reflections. In the present case, the (111) superlattice reflection is absent, whereas the (200) reflection is clearly visible, as shown in the inset of \hyperref[fig1]{Fig.~1(a)}. This observation is indicative of B2-type disorder, which is commonly encountered in QHAs and arises from the random occupation of Fe and Ru atoms as well as Mn and Ge atoms over their respective crystallographic sites. To further understand the disorder, PXRD patterns were simulated using the same atomic configuration for different degrees of B2 disorder using VESTA software \cite{VESTA}. It is observed that the simulated pattern with 25\% B2 disorder still retains the (111) superlattice reflection, which differs from the experimental observation. Interestingly, for 50\% B2 disorder, the (111) peak nearly disappears while the (200) reflection remains visible, closely matching the experimental PXRD pattern. This suggests that FeRuMnGe possesses a substantial degree of B2-type disorder involving Fe--Ru and Mn--Ge site intermixing. However, the exact disorder fraction cannot be determined solely from conventional XRD measurements, since Ge belongs to the same period of the periodic table as the transition-metal elements and therefore possesses a comparable X-ray scattering factor \cite{FeRuMnGa}.

Next, the elemental composition of FeRuMnGe specimen was examined using energy-dispersive X-ray spectroscopy (EDX). The results (as shown in \hyperref[fig1]{Fig. 1(d)}) confirm that the sample is nearly stoichiometric, with a marginal excess of Mn that lies within the permissible experimental error.

\subsection{Electronic Structure}

\subsubsection{Crystal and Electronic Structure Analysis}

\begin{figure*}
\begin{center}
    \includegraphics[width=1\linewidth]{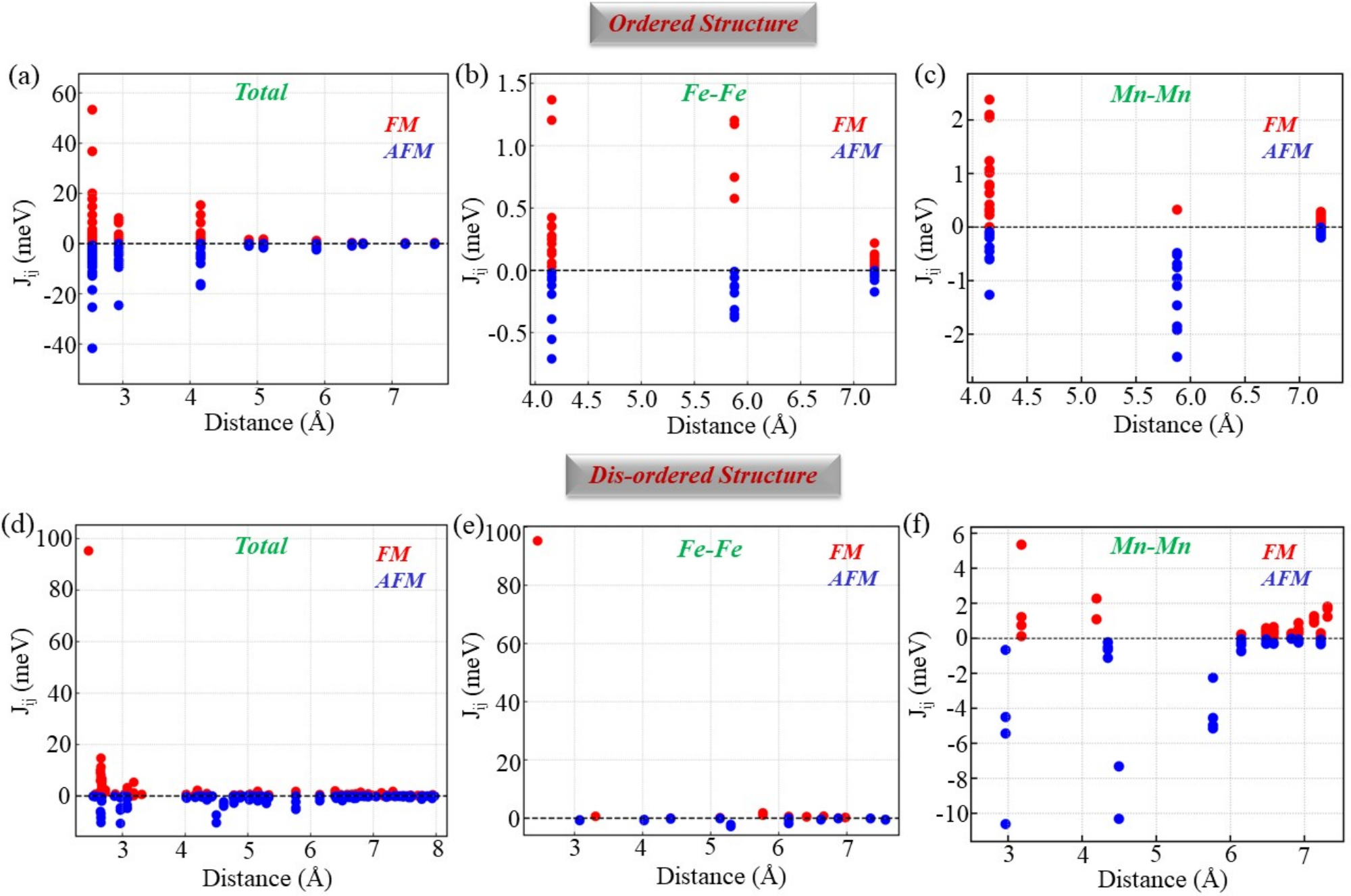}
    \caption{Heisenberg exchange interactions ($J_{ij}$) as a function of interatomic distance. Panels (a–c) correspond to the ordered structure, showing (a) total exchange interactions, (b) Fe–Fe interactions, and (c) Mn–Mn interactions. Panels (d–f) represent the Fe–Ru and Mn–Ge antisite-disordered structures, displaying (d) total exchange interactions, (e) Fe–Fe interactions, and (f) Mn–Mn interactions. Positive (negative) values of $J_{ij}$ indicate ferromagnetic (antiferromagnetic) coupling.}
    \label{exe}
\end{center}
\end{figure*}
Density functional theory (DFT) calculations were conducted to strengthen the experimental structural analysis. 
Three ordered crystal configurations were looked at, selected to precisely align with the structural models outlined in the \textcolor{blue}{Supplementary material}. Total-energy calculations reveal that the Type-YI ordered arrangement is the ground-state configuration of FeRuMnGe. Moreover, the lattice parameter for the Type-YI configuration is 5.88~\AA, which aligns remarkably well with our experimentally measured value of 5.89~\AA.

We carried out first-principles simulations to obtain a better understanding of the experimentally reported disordered structure since disorder in FeRuMnGe has been verified by the refinement of PXRD data. To investigate the effect of lattice disorder, B2 type antisite disorder of 25\% and 50\% was introduced, 
with the 50\% swapping configuration is consistent with the experimental observations. The evolution of the electronic structure from the ordered to the disordered phase is illustrated in Fig.~\ref{ba}. The spin-polarized electronic band structure and corresponding density of states (DOS) of ordered FeRuMnGe are shown in Fig.~\ref{ba}(a-c), which exhibits nearly half-metallic behavior for ordered structure, characterized by a metallic spin-up channel and a semiconducting spin-down channel. This is further supported by the DOS, which shows finite states at the Fermi level in the spin-up channel and a suppressed contribution in the spin-down channel.
In contrast, the band structure of the 50\% disordered configuration involving Fe–Ru and Mn–Ge anti-site disorder [Fig.~\ref{ba}(d-e)] reveals a fully metallic character, which is further confirmed by the corresponding DOS shown in Fig.~\ref{ba}(f). The introduction of disorder thus destroys the half-metallic nature and drives a transition to a metallic state. The atom-resolved projected density of states (PDOS) for both ordered and disordered structures are discussed in detail in the \textcolor{blue}{Supplementary Information}. These results indicate that the electronic states near the Fermi level are predominantly contributed by Mn-$d$ and Fe-$d$ orbitals.


\begin{figure*}[http]
    \centering
    \includegraphics[width=0.8\linewidth]{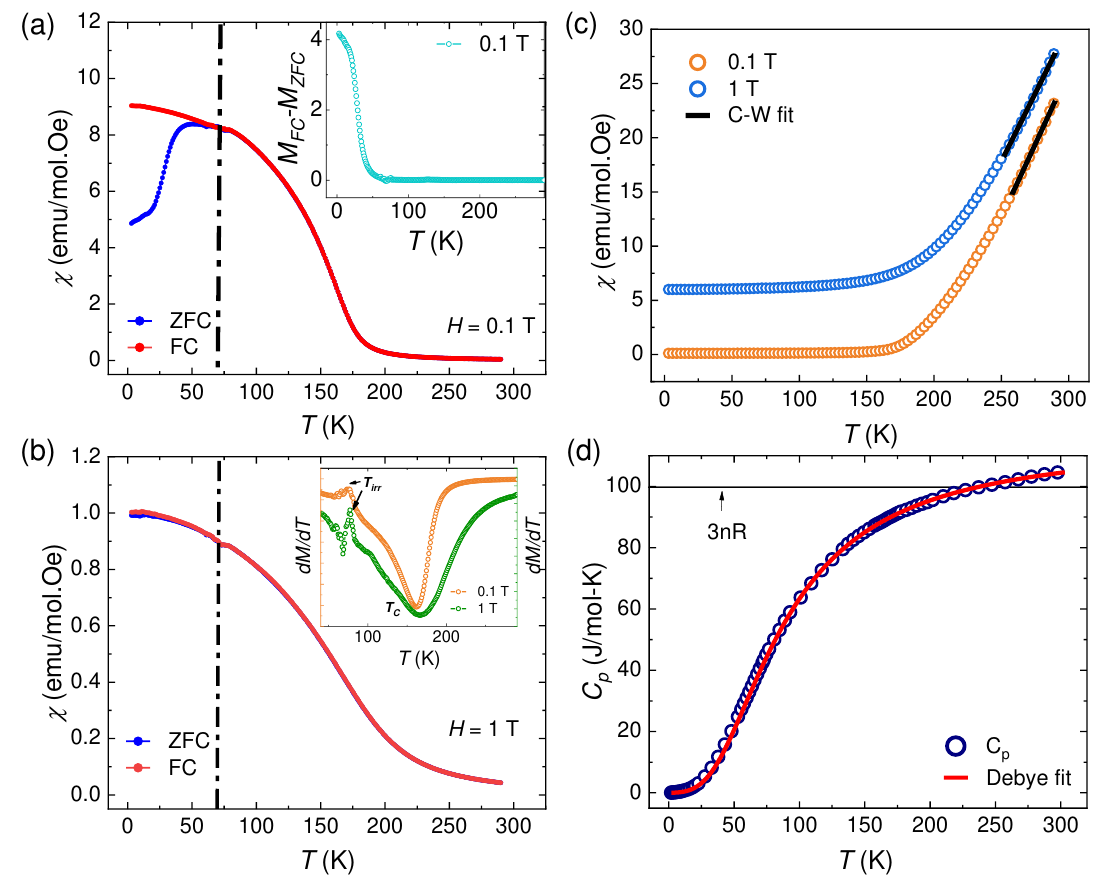}
    \caption{(Color online) (a) Temperature dependence of magnetization $\chi(T)$ measured under zero-field-cooled (ZFC) and field-cooled (FC) protocols at an applied field of 0.1 T. The inset shows $M_{\mathrm{FC}} - M_{\mathrm{ZFC}}$ as a function of temperature at 0.1 T. (vertical bar is at freezing temperature $T_f$) (b) Temperature dependence of magnetization under ZFC and FC conditions at an applied field of 1 T. (c) Inverse susceptibility $\chi^{-1}(T)$ at an applied field of 0.1 and 1 T along with the Curie--Weiss fit using Eq.~(\ref{CW}); the inset shows the corresponding $dM/dT$ vs $T$ plot at 10 kOe. (d) Inverse susceptibility $\chi^{-1}(T)$ at an applied field of 1 T fitted using the Curie--Weiss model [Eq.~(\ref{CW})]. (e) Temperature derivative of magnetization ($dM/dT$) as a function of temperature for applied fields of 0.1 and 1 T. (f) Heat capacity $C_p(T)$ of FeRuMnGe measured at zero applied field. }
    \label{fig4}
\end{figure*}
\begin{figure*}
    \centering
    \includegraphics[width=0.8\linewidth]{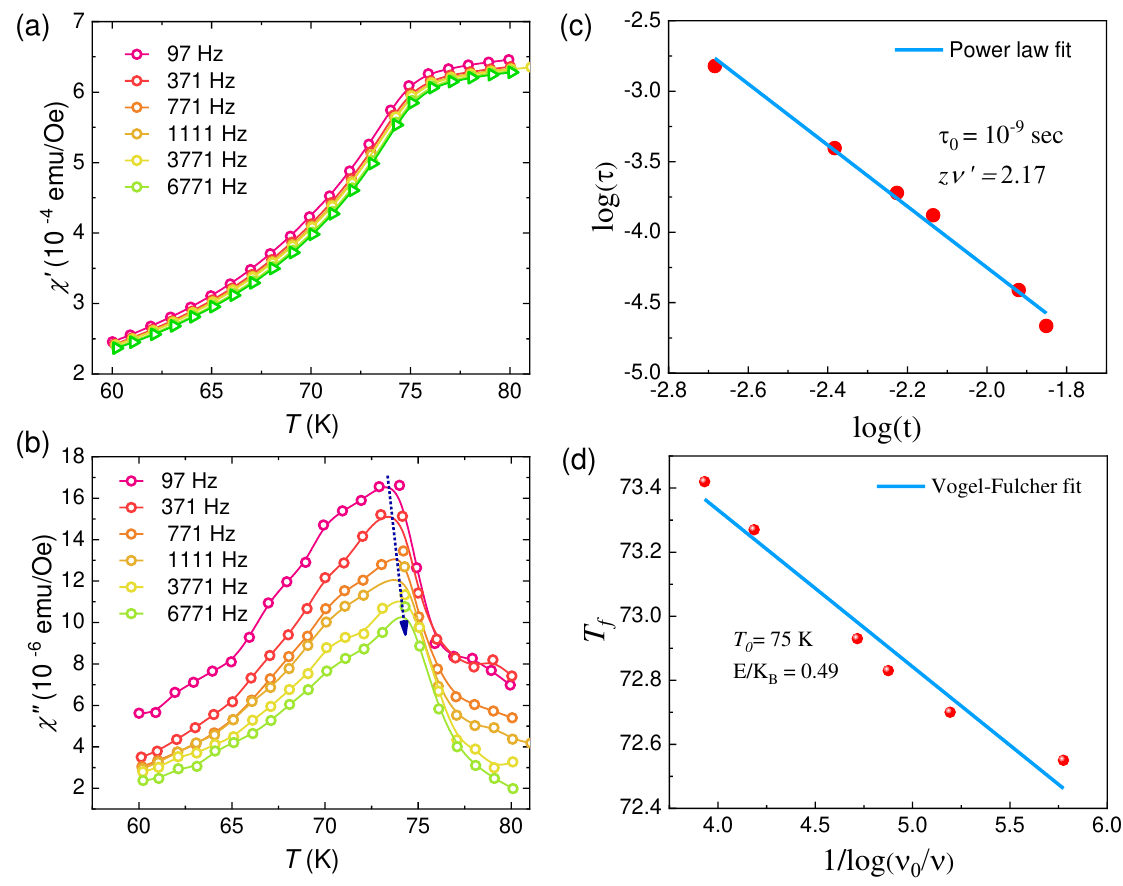}
    \caption{ AC magnetization data and fitting the shift of ${T_f}$. (a), (b) Temperature dependence of the real (${\chi'}$)and imaginary (${\chi''}$) part of the AC magnetic susceptibility measured at $H_{DC}$ = 0 Oe, $H_{AC}$= 11 Oe in the temperature range 60-80 K. (c) Critical scaling plot and (d) Vogel-Fulcher law plot.    }
    \label{fig5}
\end{figure*}

For completeness, we also examined different degrees of disorder to establish a correlation between the extent of antisite disorder and the evolution of the electronic structure. Antisite disorder between Fe and Ru, and between Fe and Mn atoms, was introduced at concentrations of 25\% and 50\% within the conventional unit cell, as detailed in \textcolor{blue}{Section~2} of the \textcolor{blue}{Supplementary Information}. The corresponding DOS reveals that increasing disorder leads to a gradual transition from a nearly half-metallic state to a fully metallic state. To investigate the physical origin of the transition from a half-metallic to a metallic magnetic state, we further analyzed the magnetic exchange interactions.

\subsubsection{Exchange Interaction}

Exchange interactions can be estimated by mapping the DFT total energies of individual spin configurations to the corresponding Heisenberg spin model \cite{Joyce1967}:
\begin{equation}
H = - \sum_{ij} J_{ij}\,\mathbf{S}_i \cdot \mathbf{S}_j,
\end{equation}

where $J_{ij}$ denotes the exchange interaction between sites $i$ and $j$, and $\mathbf{S}_i$ and $\mathbf{S}_j$ are the spin vectors at sites $i$ and $j$, respectively. The Heisenberg exchange interactions ($J_{ij}$) as a function of interatomic distance are shown in Fig.~\ref{exe}(a--f). Panels (a--c) correspond to the ordered structure, while panels (d--f) represent the disordered configuration (50 \% B2 type ).
 
For the ordered phase [Fig.~\ref{exe}(a)], the total exchange interactions exhibit competing ferromagnetic (FM) and antiferromagnetic (AFM) couplings among the first few nearest neighbors. However, the FM contributions dominate, resulting in an overall ferromagnetic behavior. The Fe--Fe interactions [Fig.~\ref{exe}(b)] are predominantly ferromagnetic at nearest-neighbor distances, indicating strong direct or hybridization-mediated exchange. In contrast, the Mn--Mn interactions [Fig.~\ref{exe}(c)] display mixed behavior, with ferromagnetic nearest-neighbor coupling and competing AFM interactions at larger distances, having similar magnitude.

In the disordered phase [Fig.~\ref{exe}(d--f)], the overall magnitude of the exchange interactions is enhanced compared to the ordered structure. The total exchange [Fig.~\ref{exe}(d)] remains predominantly positive, indicating stronger ferromagnetic ordering. The Fe--Fe interactions continue to be dominated by ferromagnetic coupling, particularly for nearest neighbors, and contribute significantly to the overall magnetism.
The oscillatory nature of $J_{ij}$ with distance, together with its gradual decay, suggests the presence of long-range exchange mediated by conduction electrons, consistent with RKKY-type interactions. Such competing interactions may introduce magnetic frustration and potentially lead to complex magnetic states. In contrast, the Mn--Mn interactions in the disordered structure [Fig.~\ref{exe}(f)] exhibit stronger competition between FM and AFM couplings. The nearest-neighbor interaction tends toward AFM, while longer-range interactions show ferromagnetic character. This enhanced competition reflects increased magnetic frustration in the disordered lattice.
To validate these theoretical findings and explore their experimental signatures, we next examine the temperature-dependent dc-susceptibility ($\chi(T)$).
\subsection{Temperature-dependent magnetic susceptibility}

\hyperref[fig4]{Fig.~4(a–c)} presents the temperature-dependent dc-susceptibility $\chi(T)$ of FeRuMnGe. The $\chi(T)$ curves measured under an applied field of 0.1~T, following zero-field-cooled (ZFC) and field-cooled (FC) protocols, are shown in \hyperref[fig4]{Fig.~4(a)}. A clear separation between the ZFC and FC curves is observed below the temperature $\simeq$ 75 K. Under ZFC conditions, upon cooling below the freezing temperature ($T_f$), the spins undergo collective freezing without any preferred orientation. When a magnetic field is subsequently applied, the spins attempt to align along the field direction; however, due to intrinsic frustration and competing interactions, the system becomes trapped in metastable states, allowing only partial alignment of magnetic moments.
In contrast, during the FC process, the presence of an external magnetic field throughout cooling imposes a directional preference, enabling the system to evolve toward a lower-energy configuration with a higher degree of spin alignment as it crosses $T_f$. As a result, the FC magnetization remains higher than the ZFC counterpart, consistent with the observed behavior. Such ZFC–FC irreversibility is a well-known signature of magnetic frustration and competing interactions (as shown in the inset of  \hyperref[fig4]{Fig.~4(a)}), and has been reported in spin-glass systems \cite{Das2018,IrMnSn, IrMnGa, NiRuMnSn}, cluster-glass materials \cite{Mukherjee1996,PANDA2026187369,paramanik2015resistivity, FeRuMnGa}, and systems exhibiting multiple coexisting magnetic phases \cite{Gabay1981}.


In systems with competing interactions, such as spin-glass materials, the magnetic response is strongly influenced by the applied field, with frustration effects being more prominent at lower fields and gradually suppressed at higher fields. Similar suppression of bifurcation is observed in our case for increase in field from 0.1~T to 1~T. Although the FC curves exhibit ferromagnetic-like characteristics, the magnetization does not saturate down to the lowest measured temperatures for either field [see \hyperref[fig4]{Fig.~4(a,b)}]. Additionally, the FC $\chi(T)$ shows a convex curvature below $T_C$, which is often considered an indicator of competing magnetic interactions \cite{Bag2018,Gondh2021,Luo2007,Anand2012,SINGH2026174167, Paramanik2020}. These experimental indications can be seen as consequence of competing AFM and FM exchange, as supported by theoretical investigation.
The $T_C$ was estimated from the temperature derivative of magnetization ($dM/dT$), and shown in the inset of \hyperref[fig4]{Fig.~4(b)}, $T_C$ shifts toward higher temperatures with increasing applied field.
Additionally, a feature observed in $\chi(T)$ near $\sim 75$~K is distinctly visible in the $dM/dT$ curves and persists even at an applied field of 1~T.
In Heusler alloys, the Curie temperature is primarily governed by interatomic exchange interactions, which are highly sensitive to variations in electronic structure and lattice parameters. As a result, $T_C$ can be effectively tuned through external parameters such as magnetic field or pressure \cite{Sasioglu2005}.

According to the Curie–Weiss (C–W) law, the magnetic susceptibility ($\chi$) follows the relation
\begin{equation}\label{CW}
\chi(T) = \frac{C}{(T-\theta_P)},
\end{equation}
where $C$ is the Curie constant and is related to the effective magnetic moment ($\mu_{\mathrm{eff}}$) through $C = N_A \mu_{\mathrm{eff}}^2 / 3k_B$, with $N_A$ being Avogadro’s number and $k_B$ the Boltzmann constant. The Weiss temperature $\theta_P$ provides information about the dominant magnetic interactions, being positive for ferromagnetic and negative for antiferromagnetic correlations.
The inverse susceptibility $\chi^{-1}(T)$ was analyzed in the high-temperature region (260–290 K), where it follows the Curie–Weiss behavior reasonably well. However, a noticeable deviation from linearity appears below approximately 225 K, which is significantly higher than the $T_C\simeq 161$ K. Such deviations from CW model are indicative of short-range spin correlations, a characteristic feature of systems with competing interactions \cite{IrMnSn,PANDA2026187369}.

From the Curie–Weiss fitting (see \hyperref[fig4]{Fig.~4(c)}), the effective magnetic moment for an applied field of 0.1 T is obtained using
$\mu_{\mathrm{eff}} = \sqrt{\frac{3k_B C}{N_A}} = \sqrt{8C}$, where $C$ is Curie-constant. The extracted value of $\mu_{\mathrm{eff}}$ is approximately $5.45~\mu_B$/f.u., while the Weiss temperature is $\theta_P \simeq 203$ K.
The positive $\theta_P$ indicates dominant ferromagnetic interactions, while the deviation from ideal Curie–Weiss behavior and the absence of magnetization saturation further support the presence of competing magnetic interactions in the system.
For an applied field of 1~T, the $\chi^{-1}(T)$ data exhibit better agreement with the Curie--Weiss model over an extended temperature range (240--290 K), yielding $\mu_{\mathrm{eff}} \simeq 5.65~\mu_B$/f.u. and $\theta_P \simeq 196$ K. Compared with the 0.1~T data, the slight increase in $\mu_{\mathrm{eff}}$ and decrease in $\theta_P$ may reflect the suppression of low-field spin fluctuations and short-range magnetic correlations, resulting in a Curie--Weiss response that more closely represents the intrinsic paramagnetic state \cite{moriya1985spin}.


To further examine the nature of the magnetic transitions, heat-capacity measurements, $C_p(T)$, were performed, as shown in \hyperref[fig4]{Fig.~4(d)}. Although long-range magnetic ordering is evident from the dc magnetic data at $T_C \sim 161$ K, no distinct $\lambda$-type anomaly is observed in the heat capacity. This may occur when the magnetic contribution is weak and masked by the dominant lattice contribution. At the characteristic freezing temperature, $T_f \sim 71$ K, the $C_p(T)$ data likewise show no discernible anomaly, consistent with the short-range magnetic correlations associated with the glassy state. Such behavior is commonly observed in re-entrant cluster-glass and related frustrated magnetic systems, where competing interactions distribute the magnetic entropy change over a broad temperature range \cite{Sharma_2026,Gao_2012,Benka_2022}. A more detailed analysis of the heat-capacity data is provided in the \textcolor{blue}{Supplementary Material}.



\begin{figure}[http]
\includegraphics[width= 1\columnwidth,angle=0,clip=true]{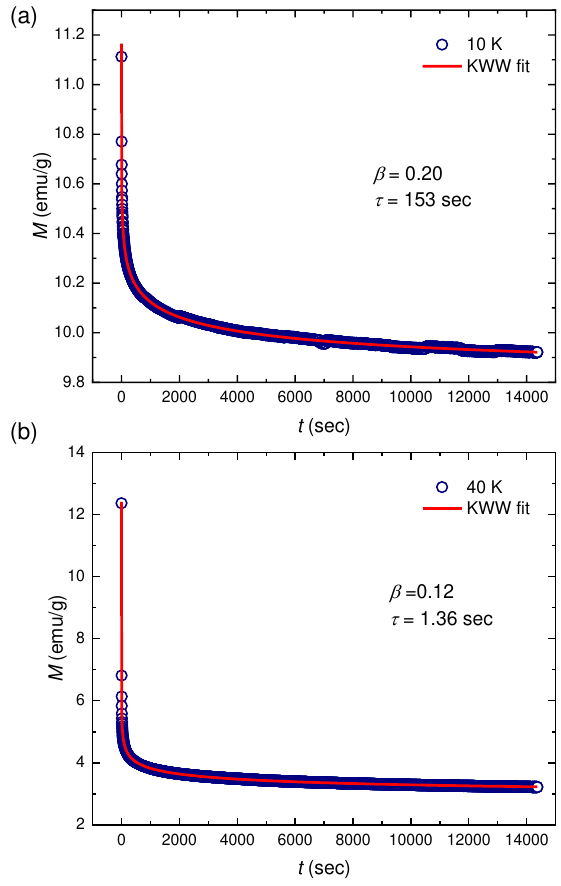}
\caption{(Color online) (a), (b) Time dependence of zero-field-cooled magnetization of FeRuMnGe measured at 10 K and 40 K, respectively, under an applied field of 10 Oe after a wait time $t_w = 10^6$ s. The solid lines represent fits to the stretched exponential function [Eq.~(\ref{TRM})].}
\label{fig6}
\end{figure}

\subsection{ac susceptibility}

The bifurcation in susceptibility data indicates spin-freezing behavior below $T_{irr}$. To further strengthen this argument, ac-susceptibility studies were conducted to investigate the fundamental magnetic dynamics \cite{Bauer2012,Topping2018}. The real ($\chi^\prime$) and imaginary ($\chi^{\prime\prime}$) components were examined across a temperature range from 50 – 85 K at various excitation frequencies (97 – 6771 Hz) employing an ac-magnetic field of 11 Oe under the ZFC method. A significant peak is noted at approximately 72.5 K, as illustrated in [\hyperref[fig5]{Fig.~5(b)}], aligning with the dc-magnetization investigations. Significantly, the peak of $\chi^{\prime\prime}$ consistently moves to a higher temperature as frequency increases, along with a decrease in the total response amplitude. The out-of-phase component ($\chi^{\prime\prime}$) emphasizes dissipative mechanisms linked to sluggish spin dynamics. This frequency-dependent behavior is a distinctive hallmark of spin-glass-like systems \cite{Pakhira2016}.

The frequency dependence of the freezing temperature ($T_f$) is further characterized by the Mydosh parameter ($S$) \cite{Mydosh1993}, which is defined as

\begin{equation}
S = \frac{\Delta T_f}{T_f(\Delta \log_{10}\nu)},
\end{equation}

where $\nu$ denotes the excitation frequency and $\Delta T_f$ signifies the variation in freezing temperature. The extent of this shift is dependent upon the magnitude of magnetic interaction: systems with strong interactions like ferromagnets or antiferromagnets have minimal frequency dependency, whereas those with lesser interactions demonstrate more significant shifts. Generally, spin glasses demonstrate $S$ values between 0.004 and 0.08, whereas superparamagnets present considerably higher values ranging from 0.3 to 0.5 \cite{Mydosh1993}. For FeRuMnGe, $S$ is determined to be 0.006 (assessed between 97 Hz and 6771 Hz), situating it at the boundary between canonical spin-glass and cluster-glass phases \cite{Mulder1981,Mulder1982,IrMnGa,Khorwal2022}.

Additionally, the frequency dependence of the freezing temperature $T_f$ was examined within the framework of dynamic scaling theory \cite{Lago2012,Malinowski2011}. The critical slowing-down behaviour of $T_f$ is described by the equation;

\begin{equation}
\tau = \tau_0\left(\frac{T_f-T_g}{T_g}\right)^{-z\nu'},
\end{equation}

where $\tau$ denotes the characteristic relaxation time linked to the excitation frequency $\nu$, and $T_g$ represents the glass transition temperature as $\nu$ approaches 0. It is widely recognized that in spin-glass systems, $T_f$ exhibits a linear relationship with $\log(\nu)$, and the extrapolation of the $T_f$ against $\log(\nu)$ graph results in $T_g \approx 71.7$ K. In this framework, $\tau_0$ denotes the microscopic spin-flip duration, $z$ signifies the dynamic critical exponent, and $\nu'$ pertains to the critical exponent of the correlation length $\varsigma$, which is defined as
$\varsigma = \left(\frac{T_f}{T_g}-1\right)^{-\nu'}$.
For practical implementation, the aforementioned equation can be reformulated as
\begin{equation}
\label{powerlaw2}
\log_{10}\nu= \log_{10}\nu_0 + z\nu'\log_{10}\left(\frac{T_f-T_g}{T_g}\right),
\end{equation}
where $\tau = 2\pi/\nu$. As illustrated in \hyperref[fig5]{Fig.~5(c)}, the analysis of the experimental data results in $z\nu' = 2.17 \pm 0.84$ and $\tau_0 \approx 10^{-9}$ s.
These parameters offer understanding into the characteristics of the glassy state. In canonical spin glasses, $z\nu'$ and $\tau_0$ generally fall within the intervals of 4--12 and $10^{-10}$--$10^{-13}$ s, respectively \cite{Lago2012,Malinowski2011}. Conversely, cluster-glass systems demonstrate somewhat elevated $\tau_0$ values, often spanning from $10^{-7}$ to $10^{-10}$ s \cite{Kumar2021, Tapas_Dy5Pd2J}. The derived $\tau_0 \simeq 10^{-9}$ s thus signifies the existence of interacting spin clusters and implies considerable magnetic clustering within the system.

In systems featuring interacting spins, the dynamics are determined by an activation energy barrier that distinguishes various magnetic configurations, which is contingent upon the quantity of collectively interacting spins and reaches its minimum in weakly correlated, paramagnetic-like states. The spin dynamics can be defined by means of the Arrhenius equation;

\begin{equation}
\tau= \tau^* \exp\left(\frac{E_a}{k_B T_f}\right),
\end{equation}

where $\tau^*$ is comparable to $\tau_0$ in the power-law framework, and $E_a/k_B$ denotes the mean activation energy linked to transitions among metastable states. The examination is conducted by graphing $\ln \nu$ against $1/T_f$. Nevertheless, the fitting produces non-physical values for $\ln \nu$ and $E_a/k_B$, suggesting that the Arrhenius model is inapplicable in this instance (plot is not shown). This type of conduct is frequently noted in spin-glass systems \cite{IrMnGa}, where a basic thermally activated process does not adequately account for the dynamics. The Vogel--Fulcher model is extensively employed to characterize the glassy dynamics of interacting spin systems. This method takes into account the development of spin clusters and their shared relaxation dynamics. The correlation between the excitation frequency ($\nu = 1/\tau$) and the freezing temperature ($T_f$) is expressed as follows;

\begin{equation}
\label{vogel1}
\tau= \tau_0 \exp\left(\frac{E_a}{k_B(T_f-T_0)}\right),
\end{equation}

where $T_0$ denotes the Vogel--Fulcher temperature, indicative of the intensity of interactions among magnetic entities. For analytical purposes, the equation may be reformulated as;

\begin{equation}
\label{vogel3}
T_f=\left(\frac{E_a/k_B}{\ln(\nu_0/\nu)}\right) + T_0.
\end{equation}
\begin{figure*}[http]
\includegraphics[width= 1\linewidth,angle=0,clip=true]{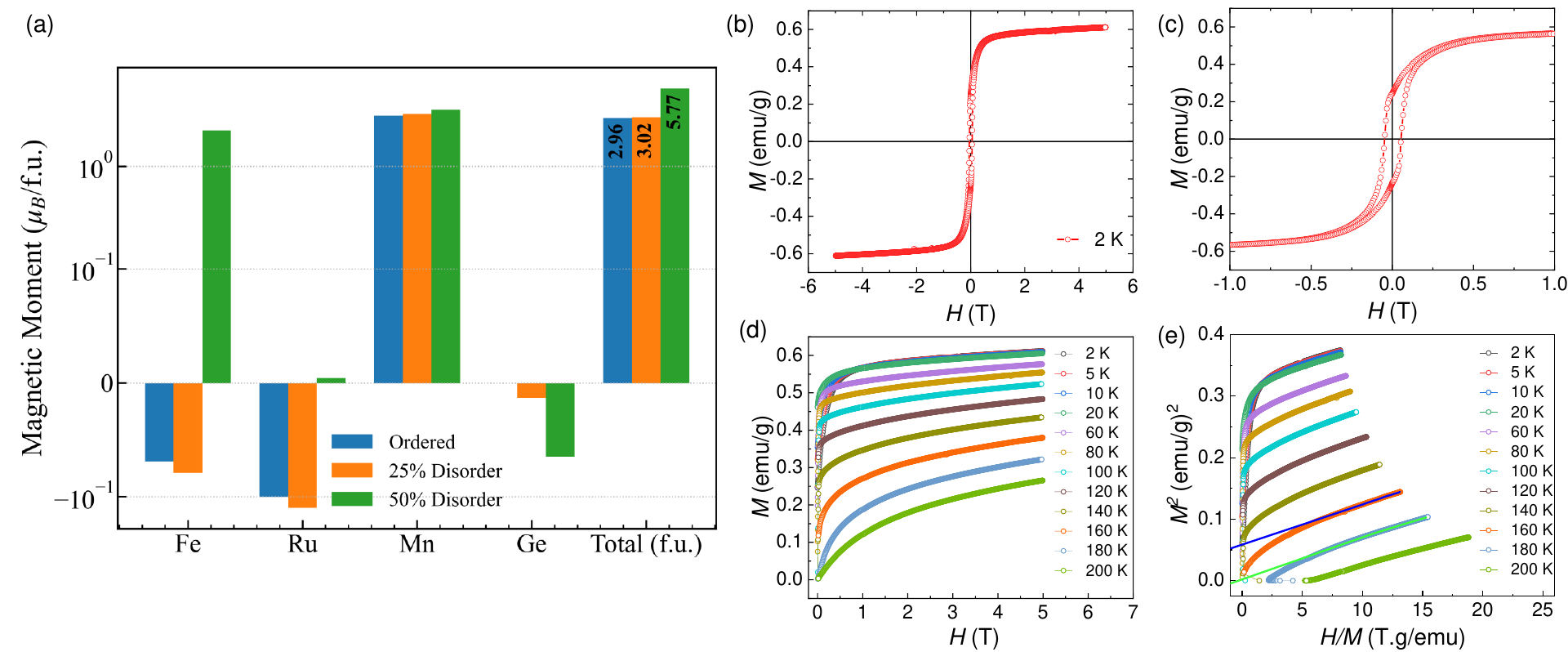}
\caption{(Color online) (a) Comparison of the average magnetic moments per formula unit ($\mu_B$/f.u.) for Fe, Ru, Mn, and Ge atoms, along with the total magnetic moment, in ordered, 25\%, and 50\% disordered FeRuMnGe structures, (b) Isothermal magnetization $M(H)$ of FeRuMnGe measured at 2 K in applied magnetic fields from $-5$ T to $5$ T.  (c) Enlarged view of $M(H)$ at 2 K in the field range $-1$ T to $1$ T. (d) Virgin magnetization curves measured at various temperatures from 2 K to 200 K. (e) Arrott plot used for the determination of the transition temperature $T_C$, obtained by extrapolating the high-field linear region, indicating a second-order magnetic transition.}
\label{MH}
\end{figure*}

\begin{figure*}[http]
\includegraphics[width= 1\linewidth,angle=0,clip=true]{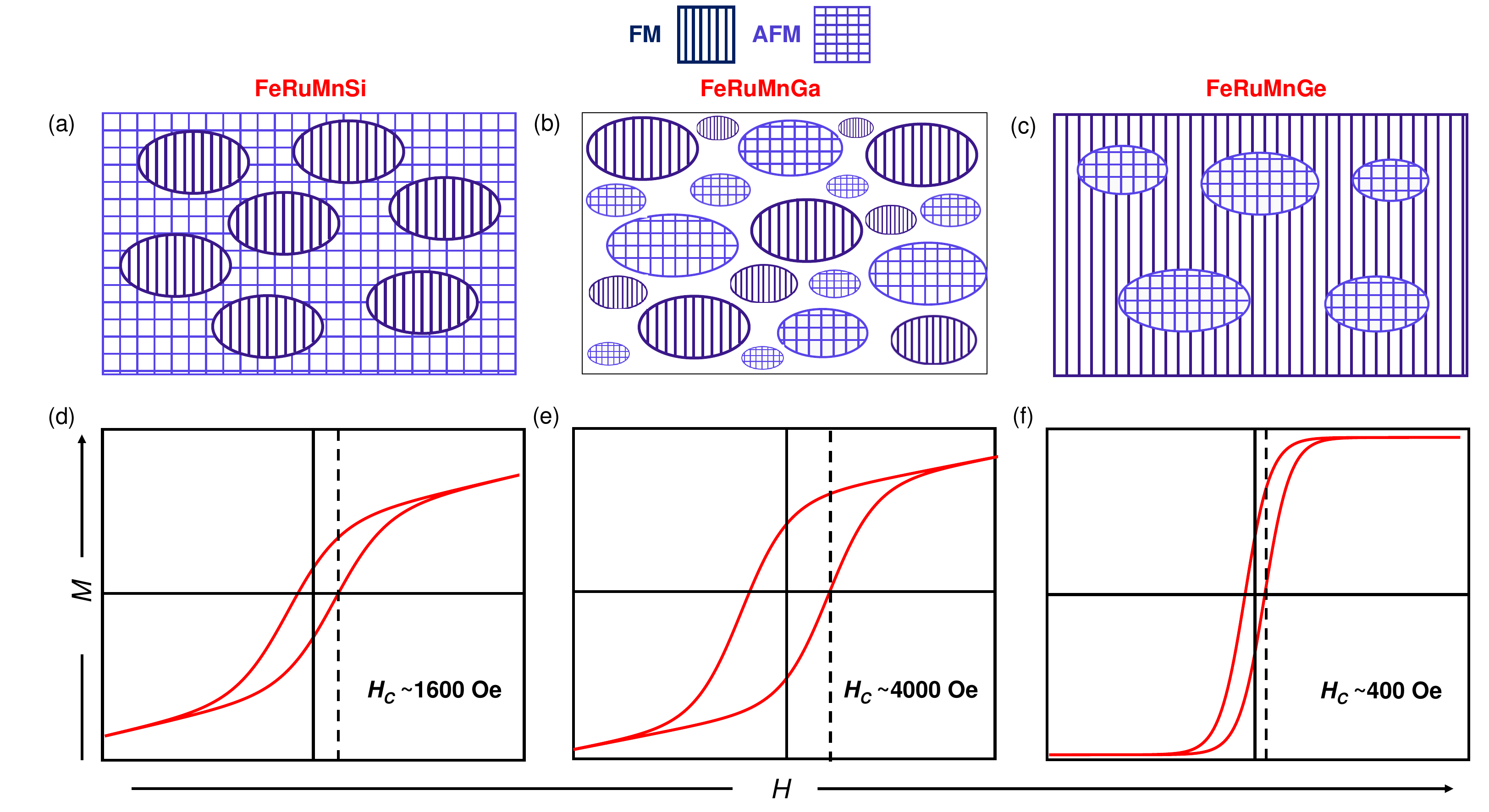}
\caption{(Color online) (a-c) Schematic of magnetic phases in FeRuMnSi, FeRuMnGa and FeRuMnGe, respectively. (d-f) Full M (H) loop for FeRuMnSi, FeRuMnGa and FeRuMnGe, respectively.  }
\label{COMPARE}
\end{figure*}

A linear dependence of $T_f$ on $1/\ln(\nu_0/\nu)$ is obtained using the Vogel–Fulcher relation, where $\nu_0$ is taken from the $\tau_0$ value extracted from the power-law analysis. From the linear fit, the parameters are found to be $E_a/k_B = 0.49 \pm 0.07$ K and $T_0 = 75 \pm 0.6$ K. These parameters provide insight into the interaction strength among magnetic entities: $T_0 \ll E_a/k_B$ corresponds to weak coupling, whereas $T_0 \gg E_a/k_B$ indicates strong interactions. In the present case, $T_0$ is approximately $\sim 150E_a/k_B$, placing the system in the strong interaction regime and suggesting significant coupling among spin clusters \cite{VFlaw}.
Further support is obtained from the Tholence criterion, $(T_f - T_0)/T_0$, which serves as a measure of glassiness \cite{Tholence1984}. The calculated value of $\simeq 0.002$ is consistent with those reported for cluster-glass systems \cite{Tholence1984,Bag2018}. Based on this comprehensive analysis, FeRuMnGe can be classified as a cluster glass system.

\subsection{Magnetic relaxation}

To investigate the glassy behavior, isothermal remanent magnetization (IRM) decay was measured at 10 K and 40 K over a duration of $10^6$~s at $T<T_C$, as shown in [\hyperref[fig6]{Fig.~6(a,b)}]. The sample was zero-field cooled from 300~K to the respective temperatures of 10~K and 40~K, followed by a wait time of $10^6$ s ($t_w$), after which a magnetic field of 10~Oe was applied. The resulting time-dependent magnetization, $M(t)$, was recorded and analyzed using a stretched exponential function,

\begin{equation}\label{TRM}
M(t)= M_0 + M_r\exp\left[-\left(\frac{t}{\tau_r}\right)^{\beta}\right],
\end{equation}

where $M_0$ denotes the intrinsic magnetization, $M_r$ represents the relaxing (glassy) component, $\tau_r$ is the characteristic relaxation time, and $\beta$ ($0<\beta\leq1$) is the stretching exponent. The parameter $\beta$ characterizes the distribution of relaxation processes \cite{Bag2018}; $\beta = 1$ corresponds to a single relaxation time, while lower values indicate a broad distribution associated with multiple interactions.

The fitting yields $\beta = 0.20 \pm 0.01$ and $0.12 \pm 0.01$, and $\tau_r = 153$ s and 1.36~s for 10~K and 40~K, respectively, indicating slow relaxation dynamics involving multiple metastable states. These values are consistent with those reported for glassy magnetic systems \cite{FeRuMnGa,Gondh2021}. The non-exponential nature of the relaxation, along with the strong temperature dependence of $\tau_r$, further supports the presence of magnetic frustration and multiple metastable states, consistent with glassy magnetic behavior in FeRuMnGe.

\subsection{Moment calculations and Isothermal magnetization}
Having established a competing magnetic system from both experiments and theory, we now turn to understanding the role of disorder in the magnitude of magnetization. We first emphasize on the theoretical calculation for total and element-resolved magnetic moments of Fe, Ru, Mn, and Ge. The Fig.~\ref{MH} (a) presents various contributions for both ordered and disordered structures. As evident from the figure, Mn atoms maintain robust and dominant magnetic moments across all configurations, whereas Fe undergoes a transition from weak antiparallel moments in the ordered phase to strong ferromagnetic moments upon the introduction of 50\% disorder. In contrast, Ru and Ge exhibit only small induced magnetic moments. In particular, the 50\% Fe–Ru and Mn–Ge antisite disorder leads to a significant enhancement of the total magnetic moment. This increase is primarily driven by the strengthening of Fe magnetism associated with Fe-$d$ states, along with a moderate increase in the Mn moment, consistent with experimental observations. In contrast, the total magnetic moment of the 25\% disordered configuration exhibits only a marginal increase, remaining nearly identical to that of the ordered structure.


\begin{table*}[ht]
\centering
\renewcommand{\arraystretch}{1.4}
\setlength{\tabcolsep}{5pt}

\caption{Comparison of structural, magnetic, and ac susceptibility parameters for FeRuMnX (X = Si, Ga, Ge).}

\begin{tabular}{c c c c c c c c}
\toprule
\toprule

\multicolumn{8}{c}{\textbf{dc Magnetic and Thermodynamic Parameters}}\\
\midrule

\textbf{Compound} &
\begin{tabular}{c}$a$\\(\AA)\end{tabular} &
\begin{tabular}{c}$T_{C}$/T$_N$\\(K)\end{tabular} &
\begin{tabular}{c}$\mu_{\mathrm{eff}}$\\($\mu_B$/f.u.)\end{tabular} &
\begin{tabular}{c}$\theta_P$\\(K)\end{tabular} &
$f$ &
\begin{tabular}{c}$M_{\mathrm{max}}$\\($\mu_B$/f.u.)\end{tabular} &
\textbf{Ref.} \\
\midrule

FeRuMnSi & 5.76 & 175 & 5.44 (0.05 T) & 178.4 & 1.02 & 0.33 (7 T) & \cite{PANDA2026187369} \\

FeRuMnGa & 5.93 & 41 & 4.90 (0.01 T) & 107.8 & 2.62 & 0.80 (7 T) & \cite{FeRuMnGa} \\

FeRuMnGe & 5.89 & 161 & 5.45 (0.1 T) & 203.1 & 1.26 & 1.64 (5 T) & This work \\

\midrule
\multicolumn{8}{c}{\rule{0pt}{3ex}\textbf{ac Susceptibility Parameters}}\\
\midrule

\textbf{Compound} &
\begin{tabular}{c}$T_f$\\(K)\end{tabular} &
$S$ &
\begin{tabular}{c}$\tau_0$\\(s)\end{tabular} &
$z\nu'$ &
\begin{tabular}{c}$E_a/k_B$\\(K)\end{tabular} &
\begin{tabular}{c}$T_0$\\(K)\end{tabular} &
\textbf{Ref.} \\
\midrule

FeRuMnSi & 8.5 & 0.1 & $10^{-6}$ & 2.86 & 0.95 & 10 & \cite{PANDA2026187369} \\

FeRuMnGa & 42.5 & 0.004 & $10^{-10}$ & 4.5 & 40.2 & 14.9 & \cite{FeRuMnGa} \\

FeRuMnGe & 72.5 & 0.006 & $10^{-9}$ & 2.17 & 0.49 & 75 & This work \\

\bottomrule
\bottomrule

\end{tabular}

\label{tab:comparison_combined}
\end{table*}


To validate the theoretical finding that disorder enhances the overall magnetization of the compound, we performed isothermal magnetization measurements. \hyperref[MH]{Fig.~7(b)} shows the isothermal magnetization $M(H)$ of FeRuMnGe measured at 2 K. The magnetization increases with applied field without clear saturation up to 5 T. The maximum magnetization reaches $\sim 1.64~\mu_B$/f.u. at $5$ T, which is significantly higher than that reported for the Ga- and Si-based counterparts \cite{FeRuMnGa,PANDA2026187369}. The compound exhibits a coercivity of $\sim 400$ Oe, indicating soft ferromagnetic character, and this value is relatively smaller than those observed for the Si- and Ga-based sister compounds. The absence of saturation and the finite curvature in $M(H)$ further suggest the presence of competing magnetic interactions.
For FeRuMnGe, the calculated magnetic moment for the ordered structure is $\sim 2.96\,\mu_B$/f.u., whereas the disordered configuration exhibits an enhanced value of $\sim 5.77\,\mu_B$/f.u. In contrast, the experimentally observed magnetic moment is significantly reduced, which may be attributed to the spin-glass nature of the system and the associated competing magnetic interactions.
Overall, the close agreement between experiment and theory demonstrates that disorder plays a crucial role in determining the magnetic response of FeRuMnGe.

Arrott plots ($M^2$ vs $H/M$) were employed to ascertain $T_C$ with greater accuracy, as illustrated in \hyperref[MH]{Fig.~7(e)}. The isotherm that intersects the origin is associated with $T_C$, where extrapolations for $T < T_C$ provide the spontaneous magnetization ($M_s$), and those for $T > T_C$ result in the inverse susceptibility ($\chi^{-1}$). Virgin $M$--$H$ curves further demonstrate enduring spin correlations above $T_C$, eliminating the possibility of a simple paramagnetic state. According to Banerjee's criterion \cite{Banerjee1964}, the positive slope of the Arrott plot confirms a second-order phase transition. The dominant ferromagnetic characteristics evident in the isothermal magnetization data indicate that the system approaches the theoretically stabilized 50\% disordered state.


\subsection{Comments on magnetic behavior of FeRuMnZ (Z= Si, Ga, Ge)}
In QHAs, compositional tuning plays a major role in tailoring their physical properties. In addition, other factors like grain size, variable chemical homogeneity, and internal stress induced by preparation techniques and methods, also exert a deterministic influence. Despite this fact, we have attempted to correlate observed magnetic behavior of FeRuMnX compounds, as presented follows;
\begin{enumerate}
          \item 
         In Heusler alloys, disorder plays a crucial role in controlling the magnetic properties, and the degree of disorder itself is strongly influenced by chemical substitution \cite{graf2010heusler}. Among the FeRuMnX (X = Si, Ga, Ge) compounds, FeRuMnSi is relatively more ordered, although it is not completely free from disorder. In comparison, FeRuMnGa exhibits a higher degree of disorder, while FeRuMnGe is most disordered among three. This enhanced disorder in the Ge-based compound can be attributed to the nearly comparable atomic radii of Mn(127 pm) and Ge (122 pm) \cite{Bainsla2016}, which facilitates site intermixing. Previous studies have also shown that substitution of Ga in place of Ge tends to increase the degree of ordering in Heusler alloys \cite{Varaprasad_2010}. Since disorder strongly modifies the nearest-neighbor environment and exchange interactions, chemical substitution indirectly governs the magnetic ground state in these compounds. In the case of FeRuMnGe, the increased disorder leads to a substantial enhancement of the Fe magnetic moment, while the moments associated with Mn and Ru remain nearly unchanged. 
         Overall, in all these systems, the Fe--Fe interaction primarily controls the magnetic nature of the compound. For FeRuMnSi, the Fe--Fe interaction remains predominantly AFM \cite{PANDA2026187369}, which evolves into a spin-glass state in the Ga-based compound \cite{FeRuMnGa} and further changes to FM in FeRuMnGe. These experimental observations are consistent with our theoretical results, suggest role of extent of disorder.
    \item  

   These disorder-driven modifications of the exchange interactions are further reflected in the experimentally observed magnetic properties of the FeRuMnX series. Experimentally, all three FeRuMnX compounds exhibit competing magnetic interactions, although their magnetic ground states differ significantly. FeRuMnSi shows an AFM transition ($T_N$), FeRuMnGa ($T_f$) undergoes a glassy magnetic transition, while FeRuMnGe exhibits a predominantly FM-like transition ($T_C$) in the $\chi(T)$ measurements. In systems with competing interactions, magnetic frustration ($f=\frac{\theta_P}{\theta_N}$) often develops, leading to suppression of the magnetic transition temperature and enhancement of coercivity. As summarized in Table~\ref{tab:comparison_combined}, FeRuMnGa exhibits the highest degree of frustration among the three compounds, which results in a significant reduction of its $T_C$. Correspondingly, the coercivity follows the trend $H_C^{\mathrm{Ga}} > H_C^{\mathrm{Si}} > H_C^{\mathrm{Ge}}$ (as shown in the Figure \ref{COMPARE}), indicating stronger frustration and competing interactions in the Ga-based compound. 
A systematic evolution in the maximum achieved magnetization is observed across the series, following the trend $M_{\mathrm{max}}^{\mathrm{Si}} < M_{\mathrm{max}}^{\mathrm{Ga}} < M_{\mathrm{max}}^{\mathrm{Ge}}$ (as shown in the Figure \ref{COMPARE}). The significantly enhanced magnetic response in FeRuMnGe further supports the disorder-driven enhancement of the overall magnetization in this compound. 
    \item  
    Although all three systems exhibit cluster-glass behavior, the parameters obtained from the ac-$\chi(T)$ analysis provide valuable insight into the nature of cluster formation and their interactions. As discussed in the ac-$\chi(T)$ section, the two important parameters obtained from the Vogel--Fulcher model are $E_a/k_B$ and $T_0$. The parameter $E_a/k_B$ is related to the activation energy barrier and represents the effective thermal energy required for the spins or magnetic clusters to overcome the relaxation barrier and change their magnetic configuration. In contrast, $T_0$ is associated with the strength of intercluster interactions and is primarily responsible for the deviation from simple Arrhenius behavior of relaxation. 
    
   A comparison of the Vogel--Fulcher parameters reveals a clear evolution of the cluster dynamics across the FeRuMnX series, as summarized in Table~\ref{tab:comparison_combined}. FeRuMnSi and FeRuMnGe exhibit relatively small activation energies, with $E_a/k_B \simeq 0.95$ K and $0.49$ K, respectively, whereas FeRuMnGa shows a significantly larger value of $\sim 40.2$ K, indicating a comparatively larger relaxation barrier in the Ga-based compound. In contrast, the Vogel--Fulcher temperature $T_0$ remains relatively small for FeRuMnSi ($\sim 10$ K) and FeRuMnGa ($\sim 14.9$ K), but increases substantially in FeRuMnGe ($\sim 75$ K). The significantly enhanced $T_0$ in FeRuMnGe suggests much stronger intercluster interactions, whereas relatively weaker coupling among magnetic clusters is present in the Si and Ga counterparts. 

The comparatively small activation energies in FeRuMnSi and FeRuMnGe indicate relatively smaller magnetic clusters, which is consistent with their predominantly AFM and FM environments, respectively. In contrast, FeRuMnGa exhibits a larger activation barrier, suggesting the presence of comparatively larger or more complex magnetic clusters arising from the coexistence of FM and AFM regions. The intercluster interaction is weakest in FeRuMnSi due to its dominant AFM background, while it becomes strongest in FeRuMnGe because of the predominantly FM environment. FeRuMnGa represents an intermediate case, where competing FM and AFM interactions remain comparatively balanced. 
Overall, substitution within the FeRuMnX family systematically drives the exchange interaction from predominantly AFM in FeRuMnSi to FM in FeRuMnGe, with FeRuMnGa lying at the intermediate regime between these two magnetic states.
    
\end{enumerate}

\section{Conclusion}
In summary, the QHA FeRuMnGe was synthesized by arc melting followed by annealing. PXRD analysis and Rietveld refinement confirm the formation of the desired phase with B2-type antisite disorder. Motivated by the experimentally observed disorder, theoretical calculations were performed, which reveal that disorder destroys the half-metallicity of the ordered structure and modifies the exchange interactions, leading to competing FM and AFM couplings. Magnetic measurements reveal competing interactions, giving rise to a cluster-glass state coexisting with long-range magnetic order. DC magnetization measurements show clear irreversibility between the FC and ZFC curves below $\sim 75$ K, indicating the onset of spin freezing. To probe the glassy dynamics, AC susceptibility measurements were carried out, where a frequency-dependent peak is observed around $\sim 72.5$ K in the out-of-phase component. Analysis using the critical power law and Vogel--Fulcher model yields a relaxation time of $\sim 10^{-9}$ s, confirming the cluster-glass nature of FeRuMnGe with strong inter-cluster interactions. The glassy behavior is further corroborated by magnetic relaxation measurements.
Overall, this work establishes the crucial role of disorder in governing the exchange interactions and magnetic ground state in QHAs.

\section*{ACKNOWLEDGMENTS}

M.P. would like to acknowledge NIT Andhra Pradesh for the fellowship. T.P. would like to acknowledge UGC DAE CSR (Grant No. CRS/2021-22/02/487) and SERB (Grant No. CRG/2022/008197) for their financial support.
S.S.P. and V.K. sincerely acknowledge the National Supercomputing Mission (NSM) for providing computational resources on ‘PARAM SEVA’ at IIT Hyderabad. S.S.P. acknowledges  DST-INSPIRE for a research fellowship and V.K. expresses gratitude for the support provided through the DRDO Project No. ERIP/ER/202312003/M/01/1853.

\bibliography{FeRuMnGe}

\end{document}


\begin{titlepage}
\centering
\renewcommand{\thefootnote}{\fnsymbol{footnote}}

\vspace*{4cm}

{\Huge\bfseries Supplementary Information\par}
\vspace{1.5cm}

{\LARGE\bfseries 
Disorder-driven competing magnetic interactions and glassy magnetic behavior in quaternary Heusler alloy FeRuMnGe
\par}
\vspace{2.5cm}
{\normalsize
\textbf{Manikantha Panda},$^{1}$

\textbf{Prabuddha Kant Mishra},$^{4}$\\
\textbf{Sonali S. Pradhan},$^{3}$

\textbf{Aarti Gautam},$^{4}$
\textbf{Bhagyashree Pol},$^{5}$
\textbf{P D Babu},$^{5}$
\textbf{Ashok Kumar Ganguli},$^{4,6}$
\textbf{V.~Kanchana},$^{3}$\footnote{E-mail: kanchana@phy.iith.ac.in}
\textbf{Tapas Paramanik}$^{1}$\footnote{E-mail: tapas.phys@nitandhra.ac.in}
\par}
\vspace{1cm}

{\small
$^{1}$Department of Physics, National Institute of Technology Andhra Pradesh,\\
Tadepalligudem 534101, India\\[0.3cm]

$^{3}$Department of Physics, Indian Institute of Technology Hyderabad,\\
Kandi 502285, Sangareddy, Telangana, India\\[0.3cm]

$^{4}$Department of Chemistry, Indian Institute of
Technology Delhi, New Delhi 110016, India\\[0.3cm]

$^{5}$UGC-DAE Consortium for Scientific Research, Mumbai Centre, BARC Campus, Mumbai 400085, India\\[0.3cm]

$^{6}$ Department of Chemical Sciences, Indian Institute of Science Education and Research, Berhampur, Odisha-760003, India\\[0.3cm]


\par}

\vfill

\end{titlepage}

\setcounter{page}{2}
\pagestyle{fancy}
\fancyhf{}
\rhead{\thepage}


\section{Crystal Structure}

\begin{figure*}
    \centering
    \includegraphics[width=0.95\linewidth]{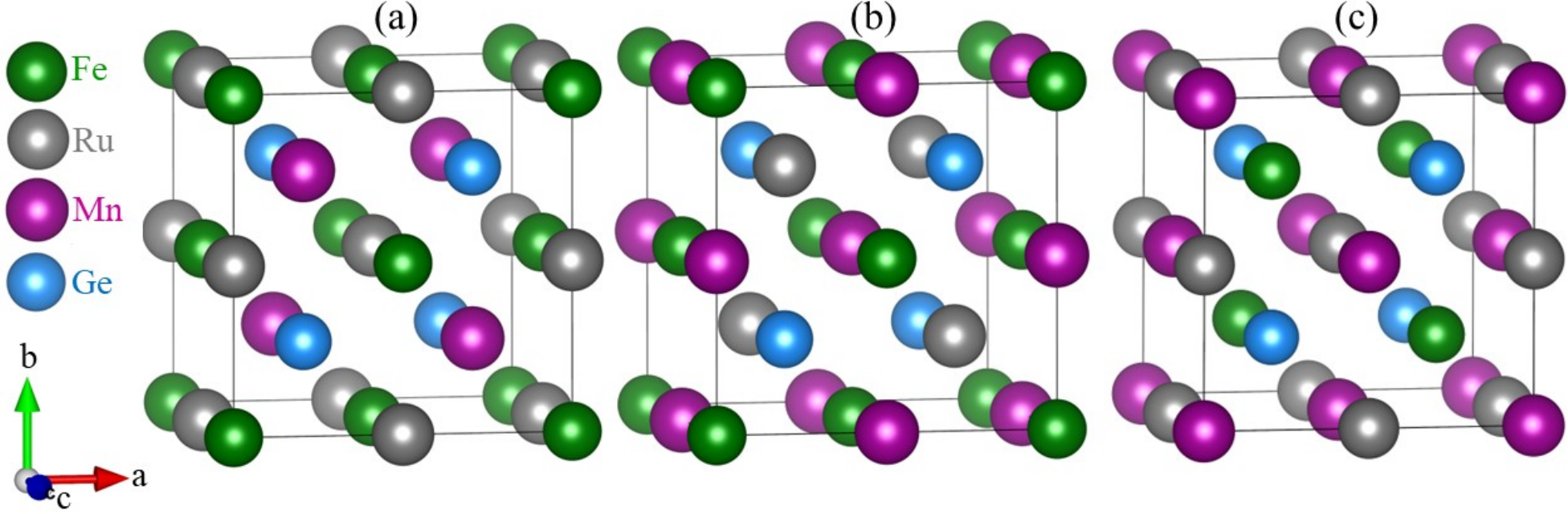}
    \caption{Crystal structure representations of the YI, YII, and YIII ordered phases of FeRuMnGe.}
    \label{cry1}
\end{figure*}

FeRuMnGe is a quaternary Heusler compound belonging to the XX$'$YZ family and adopts the LiMgPdSn-type structure with space group $F\bar{4}3m$~\cite{Bainsla2016}. Equiatomic quaternary Heusler alloys (EQHAs) comprise four inequivalent Wyckoff positions, namely $4a$~$(0, 0, 0)$, $4b$~$(0.5, 0.5, 0.5)$, $4c$~$(0.25, 0.25, 0.25)$, and $4d$~$(0.75, 0.75, 0.75)$~\cite{Neibecker2017}. In the conventional atomic arrangement, the main-group element $Z$ occupies the $4d$ site, while the transition-metal elements $X$, $X'$, and $Y$ are assigned to the $4a$, $4b$, and $4c$ sites, respectively. Depending on the atomic species occupying the $4c$ Wyckoff position $(0.25,\,0.25,\,0.25)$, three possible ordered configurations, denoted as YI, YII, and YIII, can be considered, corresponding to the occupation by $X'$, $Y$, and $X$, respectively. For FeRuMnGe, first-principles calculations show that the YI configuration possesses the lowest total energy among the considered configurations, indicating that it is the energetically preferred arrangement, as presented in \textcolor{blue}{Table~S\ref{ta1}}. Accordingly, Fe, Ru, Mn, and Ge occupy the $4a$~$(0,\,0,\,0)$, $4b$~$(0.5,\,0.5,\,0.5)$, $4c$~$(0.25,\,0.25,\,0.25)$, and $4d$~$(0.75,\,0.75,\,0.75)$ sites, respectively, as depicted in \textcolor{blue}{Fig.~S\ref{cry1}}. The calculated equilibrium lattice parameter is $a = b = c = 5.88$~\AA{}, which closely matches our experimentally determined value of $a = b = c = 5.89$~\AA{}.
 

\begin{table}[h!]
\centering
\caption{Ordered structure types, their atomic positions, and the corresponding energies.}
\label{tab:structure_energy}
\begin{tabular}{|c|p{8cm}|c|}
\hline
\textbf{Structure Type} & \textbf{Atomic Positions} & \textbf{Energy (eV)} \\
\hline
YI   & X\, (0,0,0), X$'$\,(0.5, 0.5, 0.5), Y\, (0.25, 0.25, 0.25), Z\, (0.75, 0.75, 0.75) & $-127.42844325$ \\
\hline
YII  & X\, (0,0,0), X$'$\,(0.25, 0.25, 0.25), Y\, (0.5, 0.5, 0.5), Z\, (0.75, 0.75, 0.75) & $-123.04124299$ \\
\hline
YIII & X\, (0.25, 0.25, 0.25), X$'$\,(0.5,0.5,0.5), Y\, (0, 0, 0), Z\, (0.75, 0.75, 0.75) & $-126.28666916$ \\
\hline
\end{tabular}
\label{ta1}
\end{table}

\begin{table}[htb]
\centering
\normalsize
\caption{Relaxed lattice parameter ($a_0$), atom-projected magnetic moments ($\mu_B$) of Fe, Ru, Mn, and Ge atoms, and the total magnetic moment per formula unit ($\mu_B$) for the ordered structure.}
\label{tab:frmg}
\begin{tabular}{lcccccc}
\hline\hline
Type & $a_0$ (\AA) & $m_\mathrm{Fe}$ & $m_\mathrm{Ru}$ & $m_\mathrm{Mn}$ & $m_\mathrm{Ge}$ & Total \\
     &             & ($\mu_B$) & ($\mu_B$) & ($\mu_B$) & ($\mu_B$) & ($\mu_B$/f.u.) \\
\hline
YI & 5.88 & 0.149 & -0.051 & 2.850 & -0.002 & 2.96 \\
\hline\hline
\end{tabular}
\end{table}

\begin{figure*}
    \centering
    \includegraphics[width=0.9\linewidth]{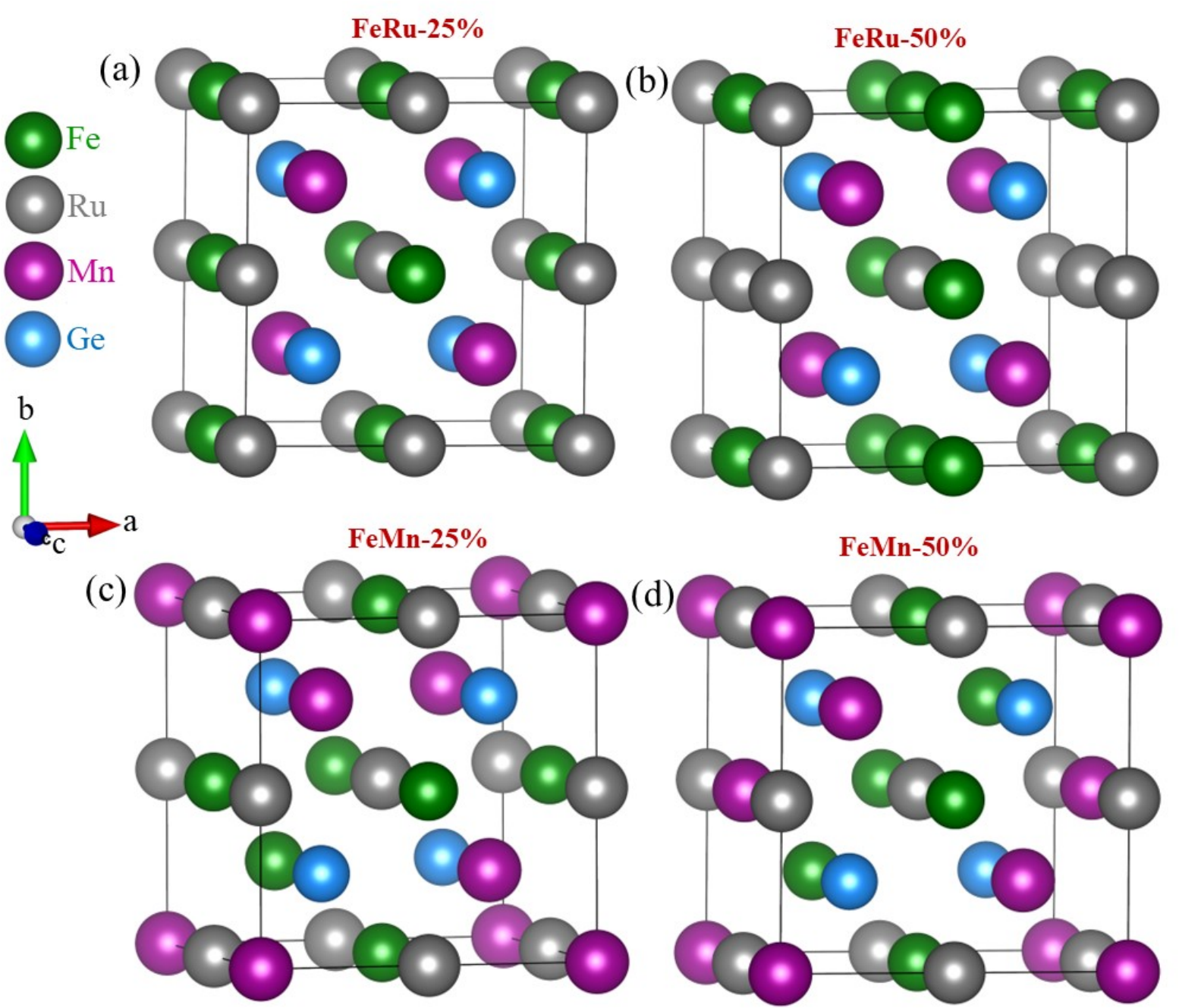}
    \caption{Fe--Ru antisite disorder at (a) 25\% and (b) 50\%, and Fe--Mn antisite disorder at (c) 25\% and (d) 50\%.}
    \label{cry}
\end{figure*}

\begin{figure*}[htb]
\centering
\includegraphics[width=0.88\linewidth]{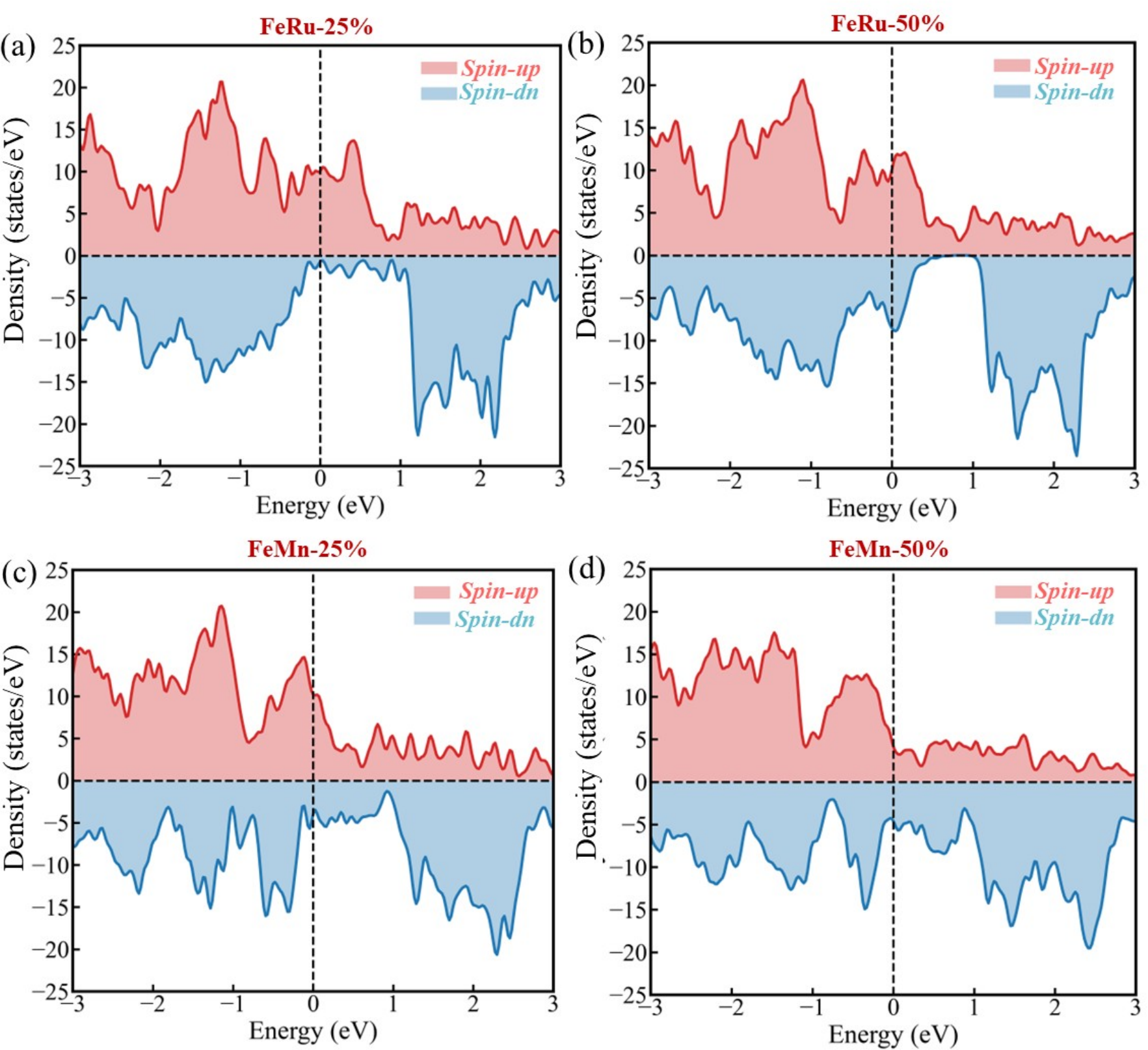}
\caption{Spin-resolved DOS of FeRuMnGe for Fe--Ru disorder [(a) 25\%, (b) 50\%] and Fe--Mn disorder [(c) 25\%, (d) 50\%].}
\label{dis_new}
\end{figure*}


\section{Disordered Density Of States}

To investigate the impact of disorder on the electronic and magnetic properties, first-principles calculations were performed using a 16-atom supercell (four formula units), where disorder was introduced via Fe--Ru and Fe--Mn atomic swaps. Owing to symmetry-equivalent sites, Fe atoms were exchanged with Ru and Mn atoms to construct different configurations. A single swap corresponds to 25\% disorder, while two swaps represent 50\% disorder. The spin-polarized density of states (DOS) for these configurations is shown in Fig.~S\ref{dis_new}(a--d). With increasing disorder, the system progressively loses its half-metallic character and evolves toward a metallic state. At 25\% disorder, the half-metallicity is weakly affected for Fe--Ru disorder but is more strongly suppressed for Fe--Mn disorder. At 50\% disorder, both cases exhibit a fully metallic behavior.



\begin{figure*}
    \centering
    \includegraphics[width=1\linewidth]{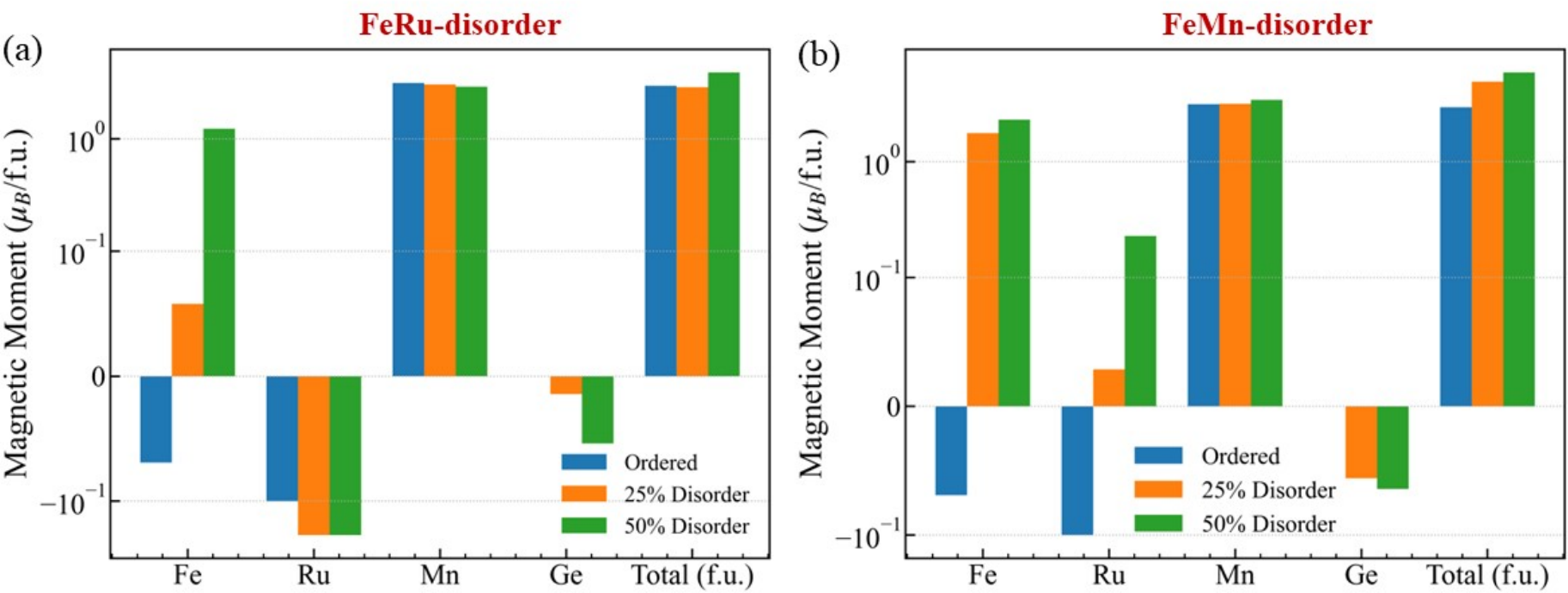}
    \caption{Comparison of the average magnetic moments per formula unit ($\mu_B$/f.u.) of Fe, Ru, Mn, and Ge atoms, along with the total magnetic moment, in ordered, 25\%, and 50\% disordered FeRuMnGe structures for (a) Fe--Ru and (b) Fe--Mn antisite disorder configurations.}
    \label{mom1}
\end{figure*}

Figure~S\ref{mom1} compares the element-resolved and total magnetic moments for Fe--Ru and Fe--Mn disorder configurations at different disorder concentrations. In the Fe--Ru disorder case [Fig.~S\ref{mom1}(a)], the introduction of disorder leads to a significant enhancement of the Fe magnetic moment, accompanied by a moderate increase in the Mn moment, while Ru retains a small negative contribution. As a result, the total magnetic moment increases substantially with increasing disorder, particularly at 50\%, indicating a disorder-driven enhancement of ferromagnetism.

In contrast, for Fe--Mn disorder [Fig.~S\ref{mom1}(b)], both Fe and Mn moments increase with disorder; however, the enhancement is comparatively more gradual. The Mn sublattice remains the dominant contributor to the total magnetization, while Ru and Ge exhibit small induced moments. Although the total magnetic moment increases with disorder in both cases, the enhancement is more pronounced for Fe--Ru disorder. This behavior highlights the crucial role of Fe in driving the magnetic response under disorder and suggests that the nature of atomic substitution strongly influences the magnetic properties of the system.

\begin{figure*}
    \centering
    \includegraphics[width=0.85\linewidth]{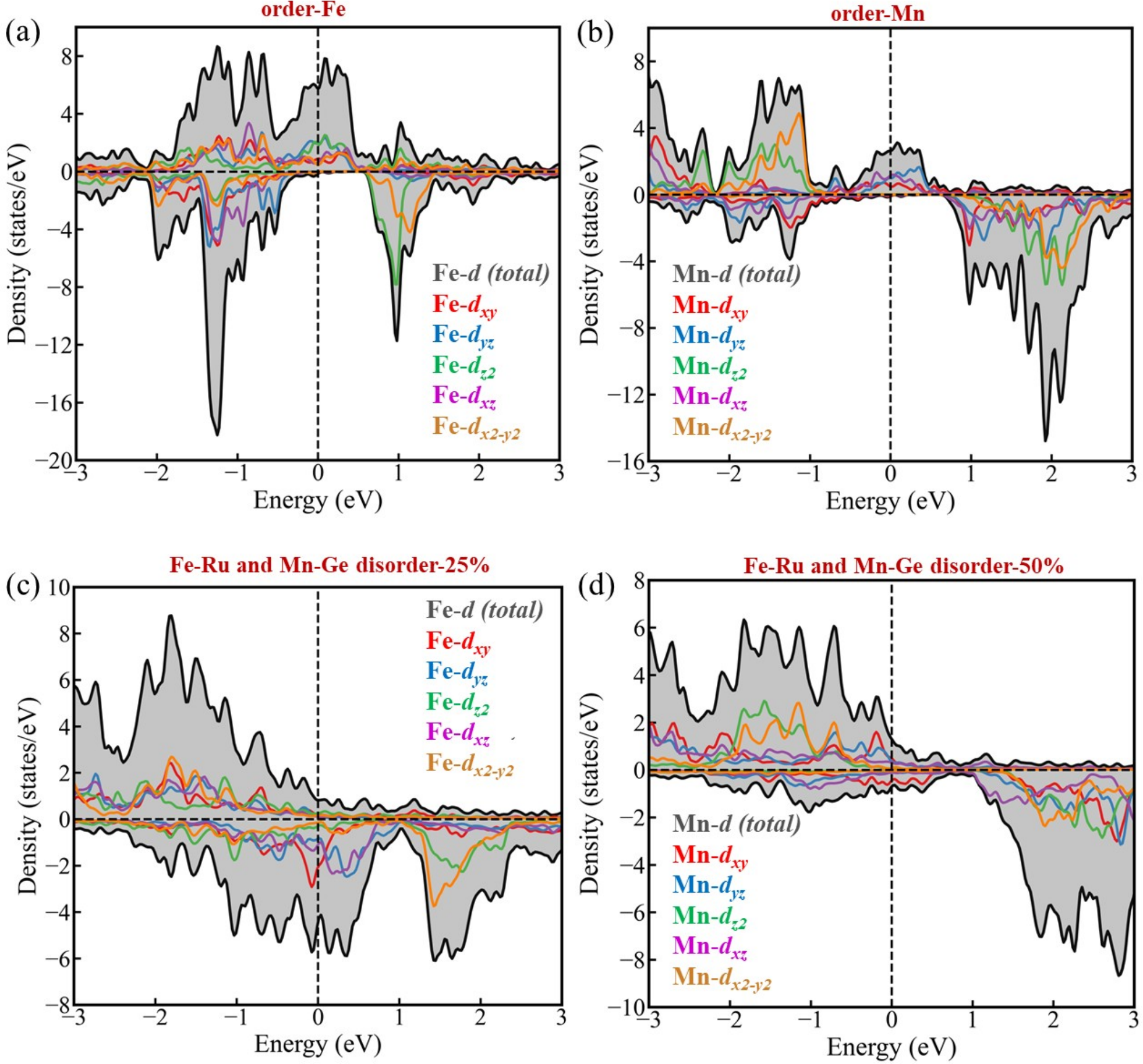}
    \caption{Orbital-projected density of states (PDOS) for Fe and Mn atoms in the ordered [(a) Fe, (b) Mn] and disordered [(c) 25\% and (d) 50\% Fe--Ru and Mn--Ge antisite configurations.]}
    \label{dis_1}
\end{figure*}



\section{Heat Capacity }

The temperature dependence of the heat capacity $C_p(T)$ of FeRuMnGe is shown in Fig.~S1. The data exhibit a smooth variation over the entire temperature range without any sharp anomaly, even near the magnetic transition temperature $T_C \sim 162$ K identified from magnetization measurements. This absence of a $\lambda$-type feature indicates that the transition is not associated with long-range magnetic ordering but rather arises from short-range correlations, consistent with spin-glass-like behavior \cite{Mydosh1993,Binder1986}.
At high temperatures, the heat capacity approaches the Dulong--Petit limit ($C_p = 3nR$), confirming the dominant contribution of lattice vibrations. The experimental data are well described by the Debye model, as shown by the solid line in Fig.~S1, yielding a Debye temperature $\Theta_D \sim 355$ K. The good agreement between the experimental data and the Debye fit suggests that the lattice contribution dominates the heat capacity in this temperature range.
In the low-temperature regime, the heat capacity was analyzed using the relation $C_p = \gamma T + \beta T^3$, where the linear and cubic terms represent electronic and phonon contributions, respectively. The obtained Sommerfeld coefficient $\gamma = 0.008$ J/mol-K$^2$ indicates a moderate electronic density of states at the Fermi level, while the $\beta$ coefficient reflects the lattice stiffness consistent with the extracted Debye temperature.

\clearpage


\bibliographystyle{unsrt}
\bibliography{FeRuMnGe}